# Hadron Spectrum and Matrix Elements in QCD with Dynamical Wilson Fermions at $6/g^2 = 5.3$

Khalil M. Bitar,[1] T. DeGrand,[2] R. Edwards,[1]
Steven Gottlieb,[3] U. M. Heller,[1] A. D. Kennedy,[1]
J. B. Kogut,[4] A. Krasnitz,[5] W. Liu,[6]
Michael C. Ogilvie,[7] R. L. Renken,[8] Pietro Rossi,[6]
D. K. Sinclair,[9] R. L. Sugar,[10]
D. Toussaint,[11] K. C. Wang[12]

[1] *SCRI, Florida State University, Tallahassee, FL 32306-4052, USA*
[2] *University of Colorado, Boulder, CO 80309, USA*
[3] *Indiana University, Bloomington, IN 47405, USA*
[4] *University of Illinois, Urbana, IL 61801, USA*
[5] *IPS, ETH-Zentrum, CH-8092 Zürich, Switzerland*
[6] *Thinking Machines Corporation, Cambridge, MA 02139, USA*
[7] *Washington University, St. Louis, MO 63130, USA*
[8] *University of Central Florida, Orlando, FL 32816, USA*
[9] *Argonne National Laboratory, Argonne, IL 60439, USA*
[10] *University of California, Santa Barbara, CA 93106, USA*
[11] *University of Arizona, Tucson, AZ 85721, USA*
[12] *University of New South Wales, Kensington, NSW 2203, Australia*

ABSTRACT: We present results of a lattice simulation of quantum chromodynamics with two degenerate flavors of dynamic Wilson fermions at $6/g^2 = 5.3$ at each of two dynamical fermion hopping parameters, $\kappa = 0.1670$ and $0.1675$, corresponding to pion masses in lattice units of about 0.45 and 0.31. The simulations include three other values of valence quark mass, in addition to the dynamical quarks. We present calculations of masses and of the decay constants of vector mesons and of pseudoscalars, including the D-meson decay constant. The effects of sea quarks on matrix elements and spectroscopy are small.

# I. INTRODUCTION

At present there are two popular ways of discretizing the Dirac operator and action on a lattice. Staggered fermions have a $U(1) \times U(1)$ chiral symmetry which protects massless quarks. Spin components are spread over several sites of the lattice, so that the number of fermion degrees of freedom per site is low, and so the bulk of numerical simulations of QCD performed to date use staggered fermions. However, the spin/flavor assignments for staggered fermions are really only valid in the continuum limit. For Wilson fermions chiral symmetry is explicitly broken and its recovery requires fine tuning. On the other hand, spin flavor assignments on the lattice are exactly as in the continuum. An exact algorithm for staggered fermions requires a multiple of four degenerate flavors of quarks, while an exact algorithm for Wilson fermions requires a multiple of two degenerate flavors. Of course, in the continuum limit, both formulations should lead to identical physics. It is therefore important to check whether this really holds.

To date, most simulations with dynamical fermions use staggered fermions, and at the lightest quark masses the ratio $m_\pi/m_\rho \simeq 0.4$ and $m_\pi > 0.20$ in lattice units. Published simulations with Wilson fermions only have $m_\pi/m_\rho \simeq 0.7$ [1]. Here we report on a large scale simulation of QCD with two light degenerate flavors of Wilson fermions, at a gauge coupling $6/g^2 = 5.3$ at two values of the quark hopping parameter, $\kappa = 0.1670$ and $0.1675$. These simulations correspond to pion masses in lattice units of about 0.45 and 0.31, and a lattice spacing of $1/a \simeq 1500 - 1800$ MeV. We used the Hybrid Monte Carlo algorithm; the simulations ran for about 2400 and 1300 simulation time units, respectively. The lattice size was $16^3 \times 32$ sites.

Before beginning a detailed discussion we briefly display the salient results of our simulation. In Fig. 1 we present an Edinburgh plot ($m_N/m_\rho$ vs $m_\pi/m_\rho$). This figure also includes data from another simulation we performed which involved quenched Wilson fermions [2]. We quantify the magnitude of hyperfine splittings in the meson and baryon sectors by comparing the two dimensionless quantities

$$R_M = \frac{m_\rho - m_\pi}{3m_\rho + m_\pi} \tag{1.1}$$

and

$$R_B = \frac{m_\Delta - m_N}{m_\Delta + m_N}. \tag{1.2}$$

Each of these quantities is the ratio of hyperfine splitting in a multiplet divided by the center of mass of the multiplet. A plot of $R_M$ vs. $R_B$ is shown in Fig. 2.

The most phenomenologically relevant matrix elements we have measured are the decay constants of vector and pseudoscalar mesons. The vector meson decay constant is

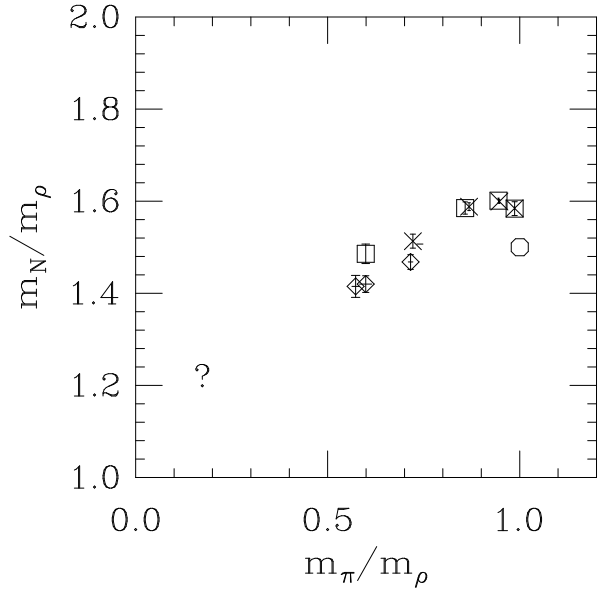

FIGURE 1

Edinburgh plot of our results. The squares are this dynamical Wilson fermion simulation at sea $\kappa = 0.1675$ and the crosses are $\kappa = 0.1670$. The diamonds are from quenched $\beta = 5.85$ and $5.95$ runs [2]. The circle and question mark show the expected values in the limit of infinite quark mass and from experiment.

parameterized by the dimensionless number $f_V$, where

$$\langle V|V_\mu|0\rangle = \frac{1}{f_V} m_V^2 \epsilon_\mu. \tag{1.3}$$

We present our calculation of $f_V$ using the lattice conserved (Wilson) vector current in Fig. 3. We see that our results show a difference of about ten per cent for the two different sea quark masses.

The second observable is the decay constant $f_P$ of a pseudoscalar meson containing one light quark and one heavy quark (such as the D or B meson). We display $f_P\sqrt{M_P}$ as a function of the inverse pseudoscalar mass $1/M_P$, since it is expected that $f_P$ scales as $1/\sqrt{M_P}$ for large $M_P$. We measured two lattice operators corresponding to the continuum axial current and converted the lattice results to the continuum using both "conventional" and "tadpole-improved" prescriptions. We show our results for each of those prescriptions in Figs. 4-5. The lattice spacing has been chosen by fitting $f_\pi$ to its real-world value, 132 MeV. Our results show little variation with respect to sea quark mass or choice of operator but considerable variation with respect to the lattice to continuum renormalization pre-

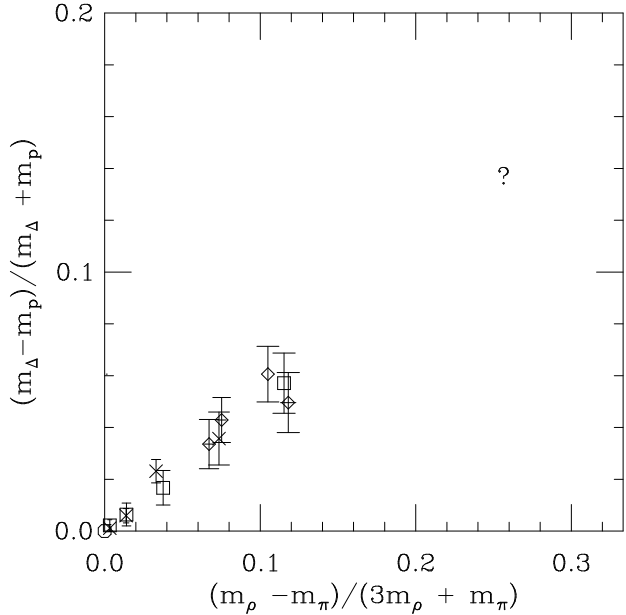

FIGURE 2

Comparison of baryon and meson hyperfine splitting, labelled as in Fig. 1.

scription we chose. Our prediction using "tadpole-improved" renormalization give about 250 MeV for $f_D$ while the "conventional" prediction is about 175 MeV. We will discuss these results and their uncertainties below.

The outline of the paper is as follows: In Section II we describe the simulations themselves. In Sec. III we review our methodology and describe our results for spectroscopy. In Sec. IV we give details of our calculations of simple matrix elements–the decay constants of vector and pseudoscalar mesons, including the decay constant of the D meson. Finally Sec. V contains some conclusions.

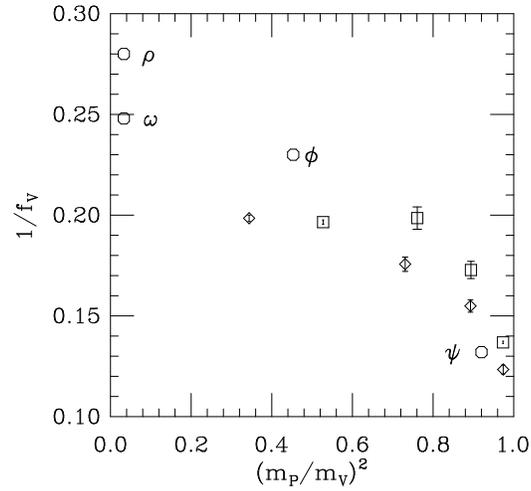

FIGURE 3

Lattice $1/f_V$ from the conserved (Wilson) vector current, as a function of the square of the pseudoscalar to vector mass ratio, $(m_P/m_V)^2$, using tadpole improved perturbation theory. The labeled points are physical particles. Results from simulations with sea quark hopping parameter $\kappa = 0.1670$ are shown in squares, and for sea quark hopping parameter $\kappa = 0.1675$ in diamonds.

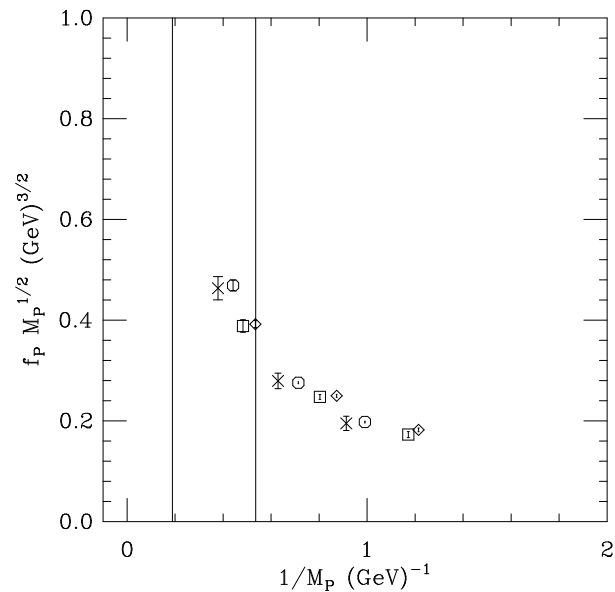

FIGURE 4

The quantity $f_P\sqrt{M_P}$ as a function of the inverse pseudoscalar mass, with lattice data analyzed using tadpole improved perturbation theory. The lattice spacing is set by fitting $f_\pi$. Our data are local and nonlocal currents at $\kappa = 0.1670$ (diamonds and octagons) and local and nonlocal currents at $\kappa = 0.1675$ (squares and crosses). The vertical lines identify the points corresponding to $f_B$ and $f_D$.

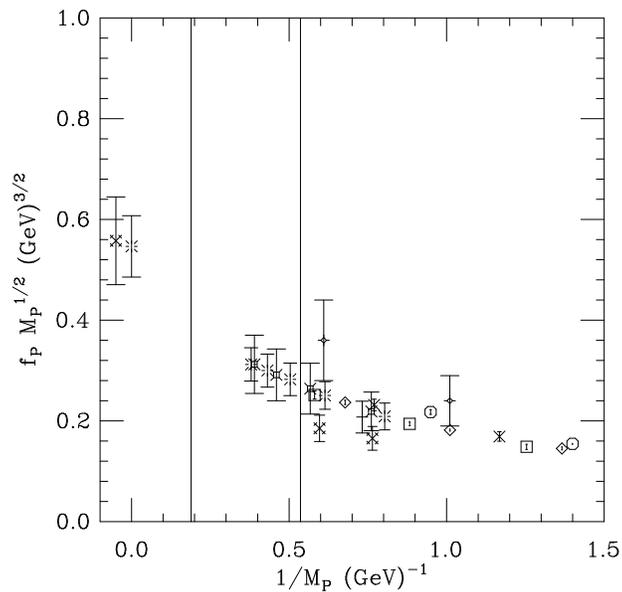

FIGURE 5

The quantity $f_P\sqrt{M_P}$ as a function of the inverse pseudoscalar mass, with lattice data analyzed using conventional field normalization. Data for static quarks are from Ref. 3 (fancy cross ), burst is Ref. 4. Other quenched heavy quark data are from the European Lattice Collaboration, Ref. 5 (fancy squares), Gavela, et. al., Ref. 6 (plus signs), and DeGrand and Loft, Ref. 7 (fancy diamonds). The scale is set by $f_\pi$. Our data are local and nonlocal currents at $\kappa = 0.1670$ (diamonds and octagons) and local and nonlocal currents at $\kappa = 0.1675$ (squares and crosses). The vertical lines identify the points corresponding to $f_B$ and $f_D$.

## II. THE SIMULATIONS

The simulations were done on the CM-2 at SCRI, with a lattice size of $16^3 \times 32$ sites. We employed the Hybrid Monte Carlo algorithm [8]. For the calculation of fermion propagators we used the Conjugate Gradient (CG) algorithm preconditioned by red-black checkerboards [9] and implemented using the fast CMIS (Connection Machine Instruction Set) inverter described in Ref. [10]. The code ran at a sustained speed of about 3 Gflops on half the machine. We chose a gauge coupling $\beta = 5.3$, somewhat smaller than in typical runs with two flavors of staggered fermions, since the renormalization of the coupling for Wilson fermions is bigger and we did not want too small a lattice spacing and hence too small a physical volume. We used two values of $\kappa$, 0.1670 and 0.1675. For $\kappa = 0.1670$ we used a conjugate gradient residual of $\sqrt{R^2/S^2} = 1 \times 10^{-5}$ in the normalization conventions of Ref. [11] and, after thermalization, time steps $dt = 0.017$ for 425 trajectories, $dt = 0.02$ and finally $dt = 0.01$ for 1000 trajectories each. These choices gave acceptance rates of about 60 per cent, 45 per cent and 80 per cent respectively. For $\kappa = 0.1675$ we used a time step $dt = 0.0069$ throughout. During the warm-up we used a CG residual of $1 \times 10^{-5}$ and observed the acceptance rate drop from about 80 to $\sim 40$ per cent. We then lowered the CG residual to $3 \times 10^{-7}$, after some tests,[12] after which the acceptance rate increased to about 90 per cent. The parameters of these runs are summarized in Table I.

The time it takes to generate a trajectory varies considerably, especially at the larger $\kappa$, closer to the critical $\kappa$, since the fluctuation in the number of iterations it takes the CG algorithm to obtain convergence are rather large. For $\kappa = 0.1675$ convergence took on the average 727 CG iterations – during the trajectory we used a linearly extrapolated guess for the starting value of the CG algorithm – with a variance of 34 per cent. For $\kappa = 0.1670$ the average number of CG iterations was 165, 199 and 149, with variances of 17 per cent, 12 per cent and 14 per cent respectively for the run segments with $dt = 0.017$, 0.02 and 0.01. The number is smallest for the smallest time step since there the extrapolated starting guess is best. The large fluctuations in the number of CG iterations required for Wilson fermions is in drastic contrast to simulations with staggered fermions. For our staggered fermion run at $\beta = 5.6$ and $ma = 0.01$ the fluctuations were about 1 per cent even though the pion mass, in lattice units, was somewhat lower. We speculate that the large fluctuations for Wilson fermions are due to the lack of a protected chiral limit. The "effective critical $\kappa$" can vary from configuration to configuration and cause these large fluctuations. This might well be the main reason why simulations with Wilson fermions appear much harder than those with staggered quarks. On half of the CM-2 it took, on average, about 4 1/3 hours to create one trajectory.

As an illustration of the time history of the runs, we display in Fig. 6 a time history of the pion propagator at separation 8, for the $\kappa = 0.1670$ run. In Fig. 7 we show the same

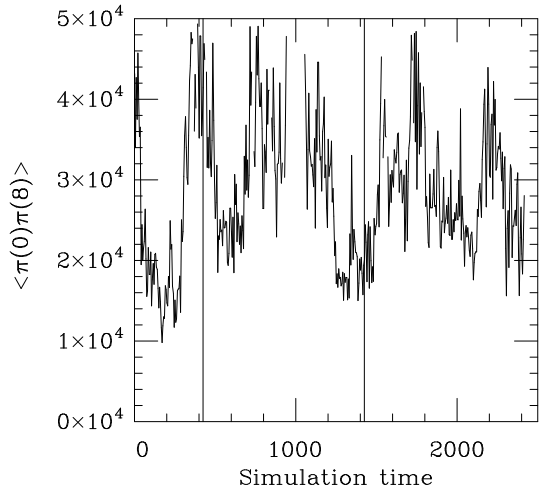

FIGURE 6

Time history of the pion propagator at distance eight for $\kappa = 0.1670$ quarks. The three regions separated by vertical lines are the parts of the run with $\Delta t = 0.017$, $\Delta t = 0.02$, and $\Delta t = 0.01$.

history for $\kappa = 0.1675$, but this time for two propagators, one whose source is on timeslice 0 and one whose source is on timeslice 16.

For the $\kappa = 0.1670$ run we measured an average plaquette $< \text{Tr} U_p > /3 = 0.52914(28)$ with an integrated autocorrelation time of about $\tau_{int} \approx 80$. Of course, a total of 2425 trajectories is not enough to obtain a reliable estimate of the autocorrelation time. For the $\kappa = 0.1675$ run the results were $< \text{Tr} U_p > /3 = 0.53354(53)$ and $\tau_{int} \approx 120$. We also measured $< \bar{\psi}\psi >$ using a stochastic estimator. Using the naive, $\sqrt{2\kappa}$, field renormalization we obtained 0.30101(5) and 0.30021(12) respectively. With the tadpole improved, $\sqrt{1 - 3\kappa/4\kappa_c}$, field renormalization and $\kappa_c = 0.16794(2)$, as determined in section 3, we find 0.22909(4) and 0.22580(9).

Later on we shall consider some matrix elements. To connect the lattice results to the continuum certain Z-factors are needed. We determine the coupling $\alpha_s$ going in their computation from the plaquette following the suggestion of [13]. For Wilson fermions the appropriate relation reads

$$-\ln\langle \frac{1}{3}\text{Tr}U_P \rangle = 4.18879 \alpha_V(3.41/a) \left\{ 1 - (1.185 + 0.025 n_f)\alpha_V + O(\alpha_V^2) \right\}. \qquad (2.1)$$

$\alpha_V(3.41/a)$ is then run down to a scale of order $1/a$ with the two-loop $\beta$-function to be used in the tadpole improved perturbative estimations of the Z-factors. In particular, for

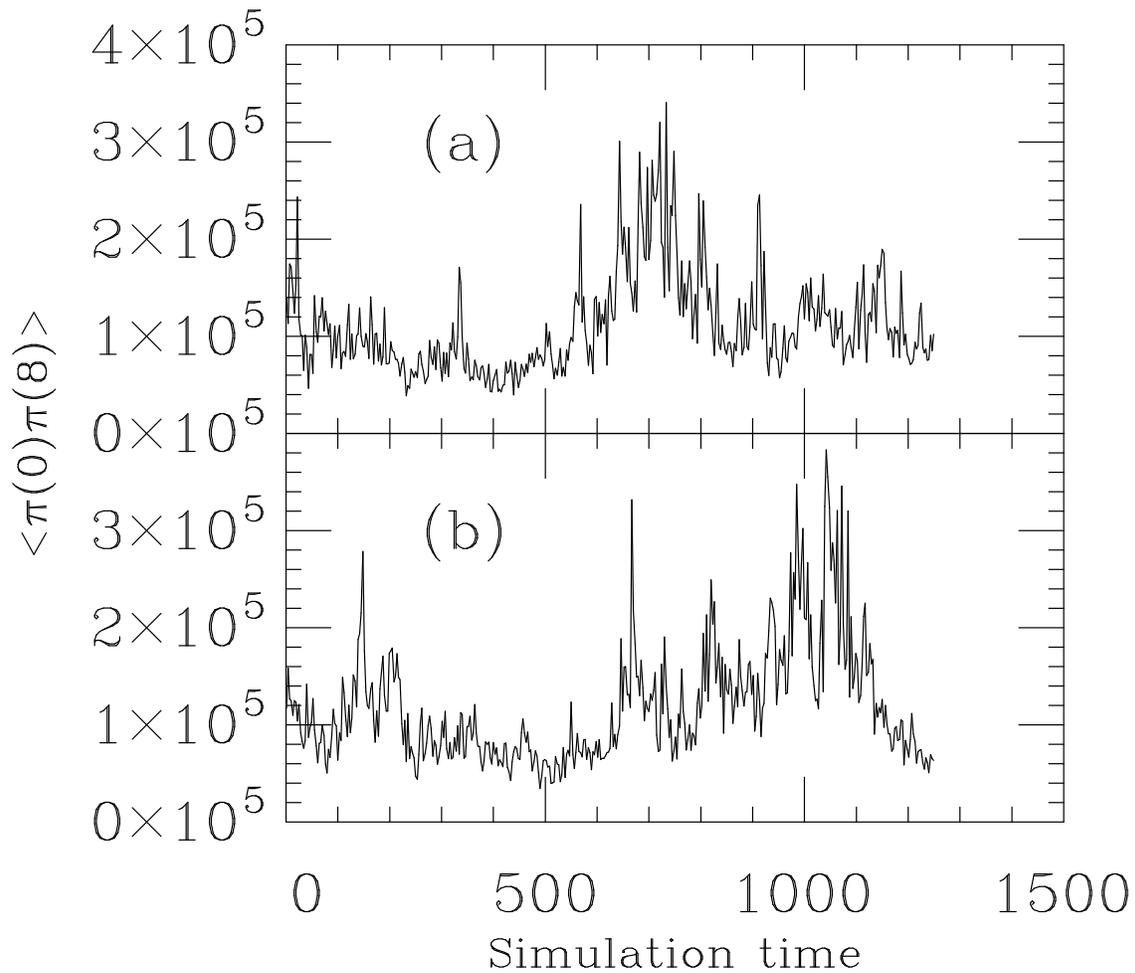

FIGURE 7

Time history of the pion propagator at distance eight for $\kappa = 0.1675$ quarks with the source at (a) $t = 0$ and (b) $t = 16$.

the tadpole improved perturbative prediction of $\kappa_c$[13]

$$\frac{1}{2\kappa_c} = 4\langle \frac{1}{3}\mathrm{Tr}\mathrm{U_P}\rangle^{1/4} - 1.268\alpha_V(1.03/a) \qquad (2.2)$$

the coupling $\alpha_V(1.03/a)$ is needed. From the measured plaquettes for $\kappa = 0.1670$ and $0.1675$ we obtain $\alpha_V(1.03/a) = 0.401$ and $0.383$. Extrapolating the plaquette linearly in $\kappa$ to $\kappa_c$, giving $<\mathrm{Tr}U_p>/3|_{\kappa_c} \simeq 0.5374$, and using this to determine the coupling, we get $\alpha_V(1.03/a) = 0.369$. Using this coupling in Eqn. (2.2) the predicted $\kappa_c$ becomes 0.1691, somewhat larger than the measured value. Were we, on the other hand, to use the relation Eqn. (2.2) and the measured values of $\kappa_c$ and the plaquette at $\kappa_c$, we would obtain $\alpha_V(1.03/a) = 0.353$. Hence, there is an uncertainty of about 0.02 in the value of $\alpha_V(1.03/a)$. Note that this uncertainty is of the same magnitude as the change that a variation of the scale, at which $\alpha_V$ is computed, by about 10 per cent would induce.

## III. SPECTROSCOPY

Our data set is summarized in Table I.

Masses and matrix elements are determined from correlation functions such as

$$C_{ij}(\vec{k}=0,t) = \sum_{\vec{x}} \langle O_i(\vec{x},t) O_j(\vec{0}, t=0) \rangle. \tag{3.1}$$

A good interpolating field is necessary so that the correlator is dominated by the lightest state in its channel at small times separation. We have chosen to fix gauge to lattice Coulomb gauge using an overrelaxation algorithm[14] and take an interpolating field which is separable in the quark coordinates and extended in the coordinates of either quark (as in the shell model):

$$O_1(\vec{x},t) = \sum_{y_1,y_2} \phi_1(\vec{y}_1 - \vec{x})\phi_2(\vec{y}_2 - \vec{x})\bar{\psi}(\vec{y}_1,t)\Gamma\psi(\vec{y}_2,t). \tag{3.2}$$

Here $\Gamma$ is an appropriate Dirac matrix, and we have suppressed all color indices. Since the operator is separable the individual $\phi$ terms are sources for calculation of quark propagators. We take $\phi(\vec{x})$ to be a Gaussian centered around the origin:

$$\phi(\vec{x}) = \exp(-(|\vec{x}|/r_0)^2). \tag{3.3}$$

The parameter $r_0$ can be chosen to give an optimal overlap with the ground state. We refer to $1/r_0 = 0$ as a "wall" source [15]; otherwise, we call the source a "shell" source [16]. At the other end of the correlator we construct either a shell sink, or a wall sink, or a point sink ($r_0 = 0$), or some matrix element, if desired.

We combine the quark propagators into hadron propagators in an entirely conventional manner. For hadrons we use relativistic wave functions [17]. For future reference baryon wave functions are:

$$\begin{aligned} &\text{Proton:} \\ &|P\rangle = (uC\gamma_5 d)u_1 \\ &\quad = (u_1 d_2 - u_2 d_1 + u_3 d_4 - u_4 d_3)u_1 \\ &\text{Delta:} \\ &|\Delta_1\rangle = (u_1 d_2 + u_2 d_1 + u_3 d_4 + u_4 d_3)u_1 \\ &|\Delta_2\rangle = (u_1 d_3 - u_2 d_4 + u_3 d_1 - u_4 d_2)u_2 \end{aligned} \tag{3.4}$$

We have measured meson correlation functions using spin structures $\bar{\psi}\gamma_5\psi$ and $\bar{\psi}\gamma_0\gamma_5\psi$ for the pseudoscalar and $\bar{\psi}\gamma_3\psi$ and $\bar{\psi}\gamma_0\gamma_3\psi$ for the vector.

We measured all hadron propagators corresponding to quarks of the same mass as the dynamical mass with wall sources and point sinks (labelled "WP" henceforth), a wall source and a wall sink (labelled "WW"), and shell sources and shell sinks or point sinks ("SS" and "SP"). We also constructed correlators for measuring masses and matrix elements with heavier valence quarks, using shell sources only. The heavier kappas are 0.1390, 0.1540, and 0.1615, plus either 0.1670 or 0.1675 (corresponding to the sea quark mass). The $r_0$ of the Gaussian function is correspondingly 2.5, 3.0, 3.5, and 4.5, and was chosen to equal the heavier ones used on the staggered sea quark analysis[18].

We measured two sets of wall source propagators on all lattices: one set is measured with the source on timeslice $t = 0$ (we call this set "kind 1") and another set with the source on timeslice $t = 16$ ("kind 2"). All data with shell sources had the source only on the $t = 0$ timeslice.

To extract masses from the hadron propagators, we average the propagators over the ensemble of gauge configurations, estimate the covariance matrix and use a fitting routine to get an estimate of the model parameters. Successive gauge configurations are not independent, so we average the propagators in blocks before estimating the covariance matrix. The block size used is discussed below in the section on results. We use the full covariance matrix in fitting the propagators in order to get a meaningful estimate of the goodness of fit. Reference 19 discusses this fitting procedure in detail.

We determined hadron masses by fitting our data under the assumption that there was a single particle in each channel. This corresponds to fitting for one decaying exponential and its periodic partner. We calculated effective masses by fitting two successive distances, and also made fits to the propagators over larger distance ranges. In addition to the use of effective masses and fits to a range of $t$ values, we show the goodness of fit of our fits to a range of $t$ by presenting pictures of "fit histograms." In these pictures a fit is represented by a rectangle centered on the best fit value for the mass, with a width given by (twice) the uncertainty of the fit (i.e. $m \pm \Delta m$), and a height which is the confidence level of the fit (to emphasize good fits) times the number of degrees of freedom (to emphasize fits over big distance ranges) divided by the statistical error on the mass (to emphasize fits with small errors). The same procedure was used in all our previous work.

We performed fits for spectroscopy from the wall sources in several ways. First, we fit "kind=1" and "kind=2" data separately to a single exponential, to see whether the masses were the same. Next, we performed a correlated fit of the two different "kind" propagators to a common mass. Then, we averaged the two "kinds" together lattice by lattice before fitting. Finally, we performed fits from shell sources and compared them to the results from wall sources. In nearly all cases good fits to a common mass were obtained. We emphasize this point because in a preliminary presentation[20] of our data the "kind=1" and "kind=2" spectroscopy gave different masses and fits to both propagators forcing a common mass had poor confidence levels. We believe that those result were due to insufficient statistics. When we quote numbers in tables from wall sources, they come from analyses in which

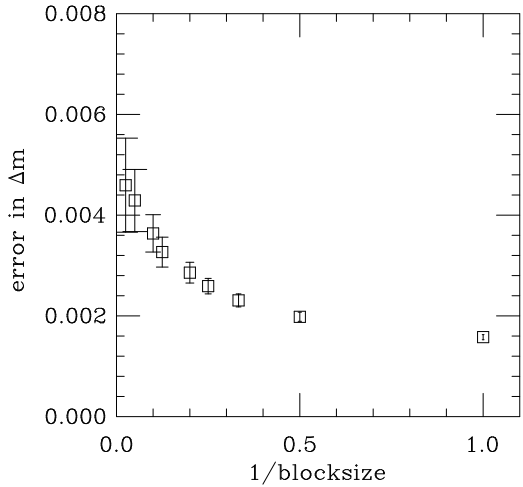

FIGURE 8

Variation in error of Wilson pion mass with block size, using the effective mass at distance 8.5 for $\kappa = 0.1670$ kind $= 1$.

the two kinds of propagators are averaged together lattice by lattice before they are fit.

We now describe the spectroscopy for each value of $\kappa$.

$$\kappa = 0.1670$$

The lattices in this data set are spaced five hybrid Monte Carlo time units apart. While a conventional autocorrelation analysis does not reveal information on the time correlations in the data set, this is probably due to the small size of our data set. We therefore looked for correlations by monitoring the variation in the uncertainty on the effective mass as we combined contiguous groups of lattices into blocks before computing spectroscopy. This reveals a long correlation time in the data. We show the variation in error of pion effective mass at distance 8.5 with block size, for $\kappa = 0.1670$, kind=1, in Fig. 8. The autocorrelation time for the pion propagator appears to be about 10 lattices or 50 time units long. We analyzed all our data by doing fits to data which had first been blocked into groups of five lattices, then into blocks of ten lattices, and extrapolated all errors to infinite blocksize with the assumption that the error varies linearly with the inverse blocksize. For the shell source data we simply blocked five lattices together (they are spaced twice as far apart in simulation time as the wall source data).

To illustrate our earlier comments about extracting masses from the various kinds of propagators, we next display a set of three fit histograms. Fig. 9 shows histograms for the kind 1 and 2 WP and WW pseudoscalars. They all appear to give the same mass. Fig. 10 shows histograms to WP particles where both kinds of propagators are fit to a common mass, and the values of the histograms are high. Finally, averaging the two kinds of propagators before fitting also produces high quality fits, as shown in Fig. 11.

The shell wave function data for mesons containing the most heavy quark ($\kappa = 0.1390$) generally have very poor fits, with a chi-squared per degree of freedom much greater than 2 or 3. We encountered this problem with heavy Wilson quarks in our earlier work [18].

While the shell and wall rho, nucleon, and delta agree within statistical uncertainties, Table II shows that the pions are many standard deviations apart (0.462(3) from the shell, 0.454(2) from the wall). We believe this discrepancy is due to lingering time correlations in our data which causes us to underestimate statistical errors.

We display plots of effective mass in Fig. 12 and of mass versus $D_{min}$ (with $D_{max} = 16$) for $\kappa = 0.1670$ data in Fig. 13. Masses from shell sources and point sinks are shown in Fig. 14. Tables II to V show our best-fit masses.

$$\kappa = 0.1675$$

In this data set the wall source lattices are spaced three Monte Carlo time units apart. The analog of Fig. 8 for $\kappa = 0.1675$ is shown in Fig. 15, and Figs. 9-14 are reproduced for this quark mass in Figs. 16-21. The data are very correlated; there does not appear to be a flattening in the uncertainty in the effective mass. This means that it is likely that our uncertainties in the fit masses are underestimated because of the correlations of the data in simulation time.

For the three parameter correlated fits to a common mass we blocked five and ten lattices together and extrapolated to infinite blocksize; for the other two fits (fitting the two kinds separately or averaging them together) we blocked ten and twenty contiguous lattices together and then extrapolated to infinite blocksize.

The shell source data are analyzed simply by blocking three successive lattices together; since they are spaced three times the wall lattices apart this is like blocking the wall lattices in groups of nine. It leaves 47 lattices to analyze. A smaller number would mean that the elements of the correlation matrix are less precisely known and could lead to singular correlation matrices.

When we compare the $\kappa = 0.1670$ and 0.1675 fits we see that the lighter quark mass data are noisier and the fits are of lower quality. The situation with the three types of fits

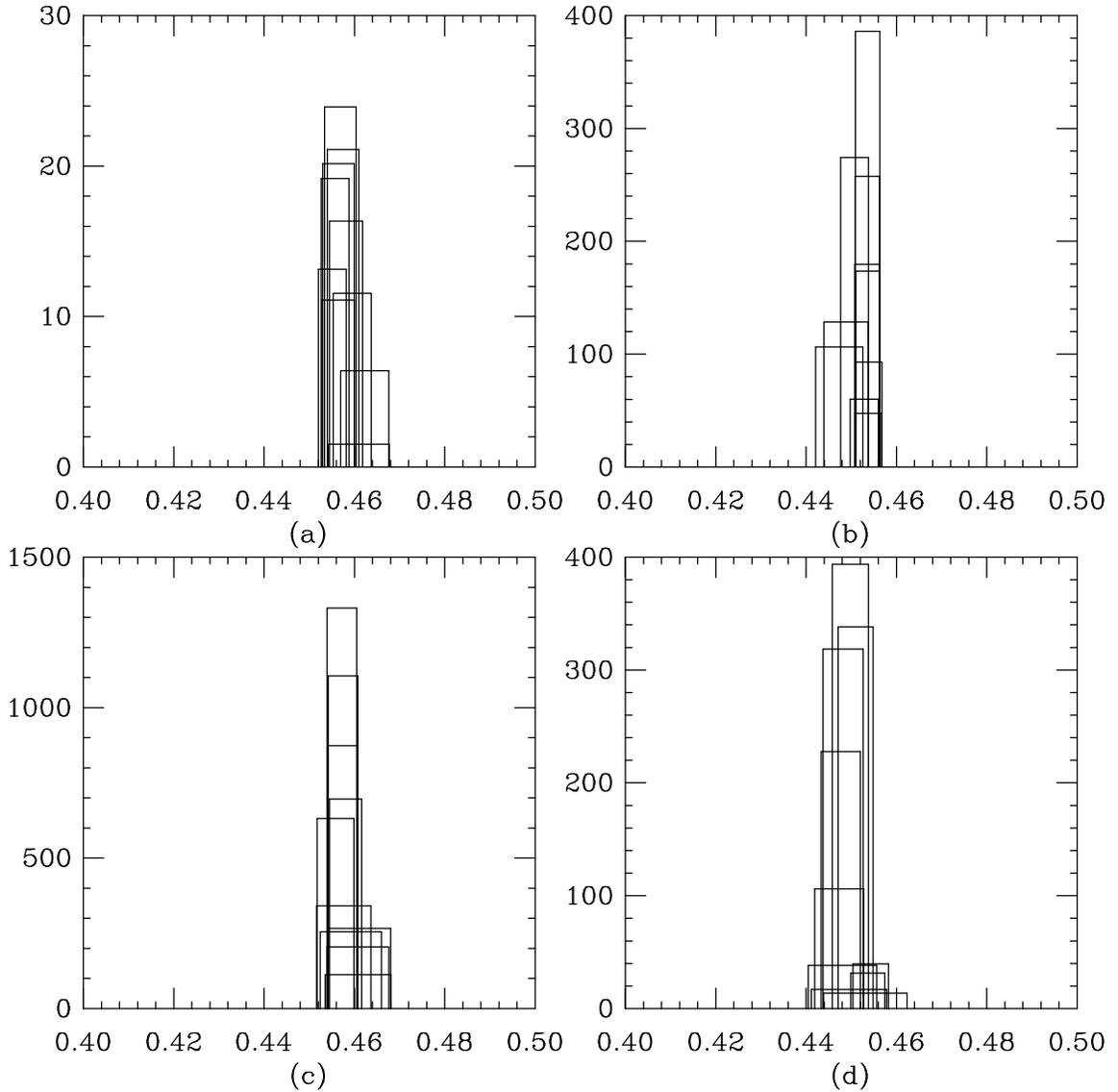

FIGURE 9

Fit histograms for pseudoscalars at $\kappa = 0.1670$ dynamical fermion data: (a) kind=1 WP correlator, (b) kind=2 WP correlator, (c) kind=1 WW correlator, and (d) kind=2 WW correlator.

(separate kinds, two kinds to a common mass, and averaged kinds) is not as satisfactory as for $\kappa = 0.1670$ data. Separate fits are shown in Fig. 16. Fits to a common mass have much lower histograms (except for the rho) and are shown in Fig. 17. Finally, averaged fits for the mesons have acceptable confidence levels, while the scatter in best fit values for

FIGURE 10

Fit histograms from correlated fits of the propagators with two different source points to a common mass for $\kappa = 0.1670$ dynamical fermion data: (a) pseudoscalar, (b) vector, (c) nucleon, and (d) delta.

the baryons is rather large. This time the shell and wall source pions' masses agree, but the statistical uncertainties are much larger than at $\kappa = 0.1670$. Results are tabulated in Tables VI-IX.

FIGURE 11

Fit histograms from fits of the propagators with data for two different source points averaged together for $\kappa = 0.1670$ dynamical fermion data: (a) pseudoscalar, (b) vector, (c) nucleon, and (d) delta.

Mass Ratios

In Fig. 1 we present an Edinburgh plot ($m_N/m_\rho$ vs $m_\pi/m_\rho$). This figure also includes data from the other simulation we performed which involved quenched Wilson fermions [2].

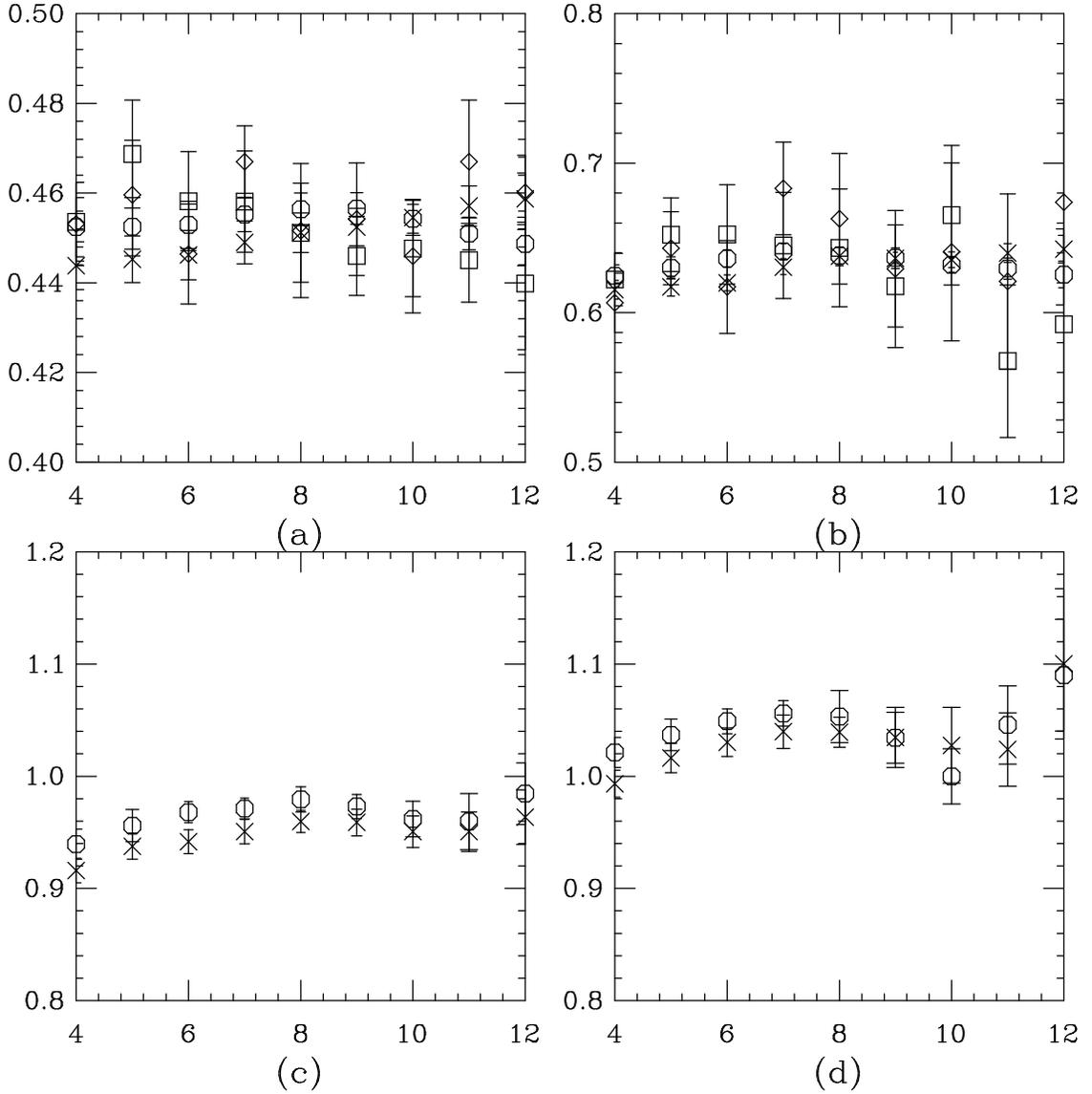

FIGURE 12

Effective mass fits to $\kappa = 0.1670$ data: (a) pion, (b) rho, (c) proton, and (d) delta. Data are labelled by type (WP or WW) and kind (1 or 2) by crosses (WP1), circles (WP2), diamonds (WW1) and squares (WW2).

There is weak evidence from this plot that the nucleon to rho mass ratio is slightly higher than from quenched simulations at $\beta = 5.85 - 5.95$ at equivalent $m_\pi/m_\rho$. In the quenched simulations the inverse lattice spacing is a little larger: 1800 to 1950 MeV vs. about 1700 MeV here. There is some evidence from quenched [21] and dynamical staggered [22]

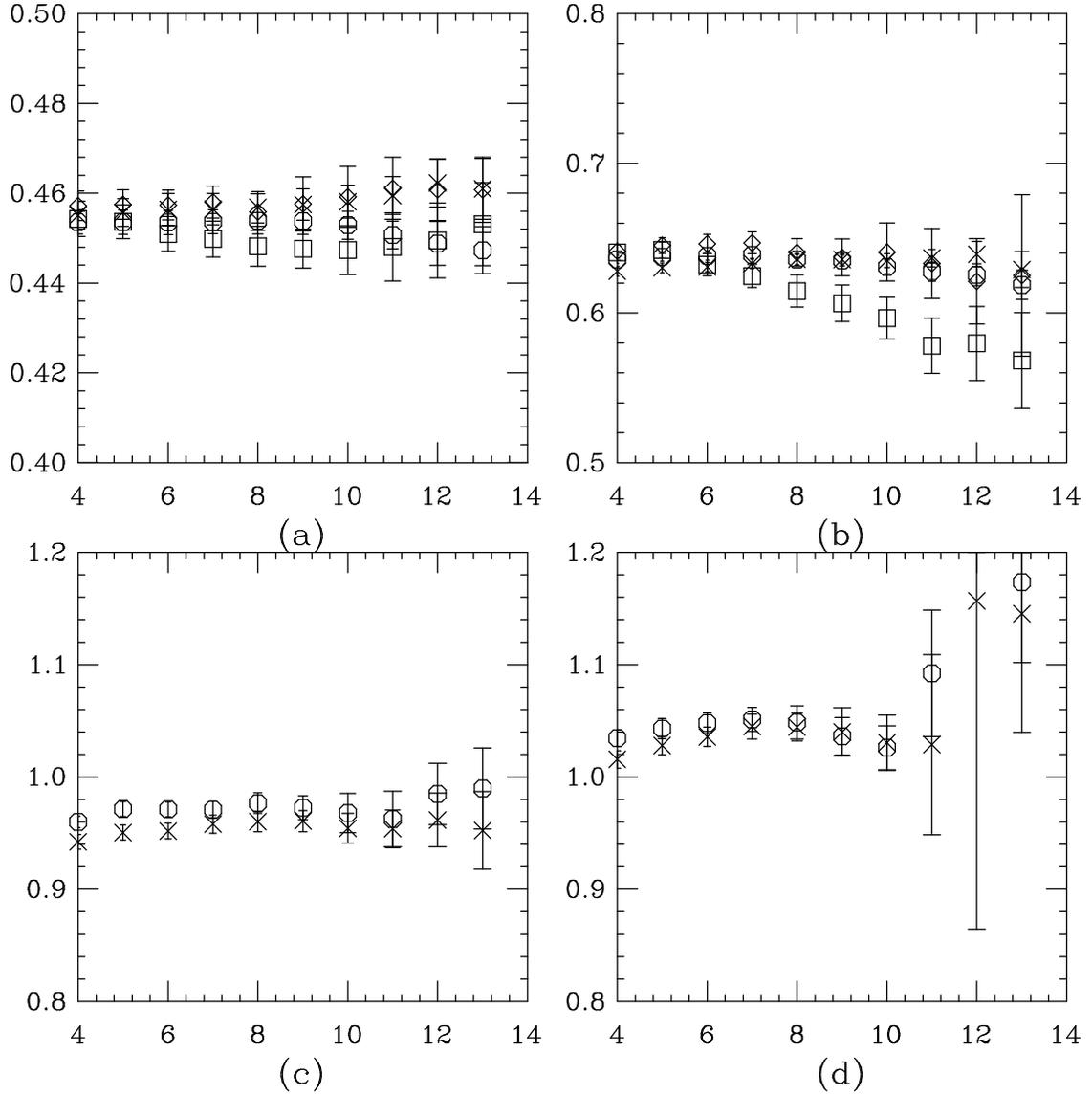

FIGURE 13

Fits from $t = D_{min}$ to 16 to $\kappa = 0.1670$ data: (a) pion, (b) rho, (c) proton, and (d) delta. Particles are labelled as in Fig. 12.

simulations that $m_N/m_\rho$ falls with decreasing lattice spacing. Fig. 22 shows a comparison of our dynamical data with that of Gupta, et. al. [23]. In this figure all valence and sea quarks are degenerate. Our results from simulations with degenerate valence and sea quarks overlaps with that of Ref. [23], but our data with heavy valence quark mass and light sea quark mass have a smaller $m_N/m_\rho$ value for large $m_\pi/m_\rho$ than the all-degenerate

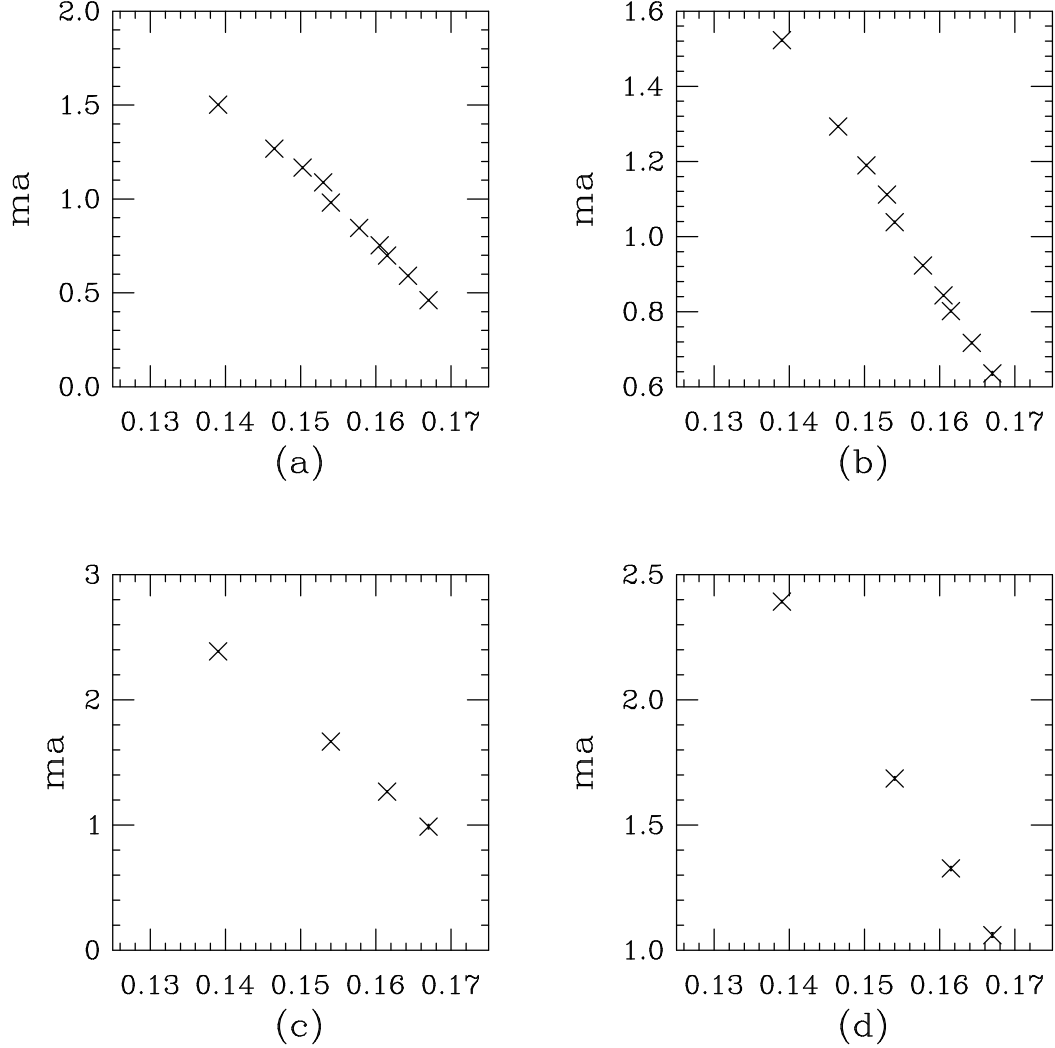

FIGURE 14

Best fit masses vs. average hopping parameter for $\kappa = 0.1670$ dynamical fermion data: (a) pion, (b) rho, (c) proton, and (d) delta.

fermion data of Ref. [23]. For these plots we performed a correlated four-parameter fit to the two mass combinations. Our best-fit values for the ratios are given in Tables X-XI.

To give an overview of our observed hyperfine splittings, we plot of $(m_\rho - m_\pi)/(3m_\rho + m_\pi)$ vs. $(m_\Delta - m_N)/(m_\Delta + m_N)$ in Fig. 2. There is nothing unexceptional in this plot.

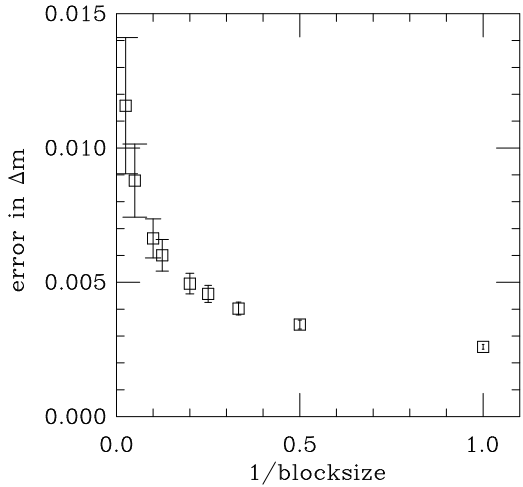

FIGURE 15

Variation in error of Wilson pion mass with block size, using the effective mass at distance 8.5 for $\kappa = 0.1675$, with a wall source and point sink.

Extrapolation to $\kappa_c$

Assuming that $m_\pi^2$ is linear in $\kappa$ (as we expect from current algebra considerations) we can compute the critical coupling $\kappa_c$ at which the pion becomes massless. There are actually three interesting critical $\kappa$'s: one is the critical coupling for two flavors of dynamical fermions at $\beta = 5.3$ and the other two are the hopping parameter values at which a pion with two valence quarks whose mass is varied while the dynamical mass is held fixed, at $\kappa = 0.1670$ or $0.1675$, becomes massless. We refer to the former $\kappa$ as the "true" $\kappa_c$ and the latter two as "fixed-background" $\kappa_c$'s. Plots of squared pion masses in fixed background are shown as a function of $1/\kappa$ in Fig. 23; we do not show a graph of pion mass squared appropriate to the true $\kappa_c$ since there are only two data points.

We look for a $\kappa_c$ in two ways. First, we extrapolate the square of the pion mass quadratically to zero via

$$(m_\pi a)^2 = A(\frac{1}{\kappa} - \frac{1}{\kappa_c}). \tag{3.5}$$

Second, we compute a quark mass from the current algebra relation

$$\nabla_\mu \cdot \langle \bar\psi \gamma_5 \psi(0) \bar\psi \gamma_5 \gamma_\mu \psi(x) \rangle = 2m_q \langle \bar\psi \gamma_5 \psi(0) \bar\psi \gamma_5 \psi(x) \rangle \tag{3.6}$$

FIGURE 16

Fit histograms for pseudoscalars at $\kappa = 0.1675$ dynamical fermion data: (a) kind=1 WP correlator, (b) kind=2 WP correlator, (c) kind=1 WW correlator, and (d) kind=2 WW correlator.

and extrapolate $m_q$ linearly to zero as in Eqn. (3.5). For the particular lattice realization of Eqn. (3.6) which we use, see the discussion in Ref. 18. We studied both the local axial current $\bar{\psi}\gamma_0\gamma_5\psi$ and the nonlocal axial current where the two fermion operators are separated by a link variable.

FIGURE 17

Fit histograms from correlated fits of the propagators with two different source points to a common mass for $\kappa = 0.1675$ dynamical fermion data: (a) pseudoscalar, (b) vector, (c) nucleon, and (d) delta.

The true $\kappa_c$ is just found by extrapolating the appropriate operators from the two data sets. The extraction of the fixed-background $\kappa_c$ is more difficult since the data are correlated and so we perform a jackknife analysis. At $\kappa = 0.1670$ we begin by blocking five contiguous lattices together before performing correlated fits, and for $\kappa = 0.1675$

FIGURE 18

Fit histograms from fits of the propagators with data for two different source points averaged together for $\kappa = 0.1675$ dynamical fermion data: (a) pseudoscalar, (b) vector, (c) nucleon, and (d) delta.

we blocked two lattices together. We then performed a jackknife dropping sets of six contiguous blocked lattices at $\kappa = 0.1670$, and at $\kappa = 0.1675$ we performed a jackknife removing 10 successive blocked lattices from the ensemble.

At $\kappa = 0.1670$ the two different procedures give two different fixed-background $\kappa_c$'s:

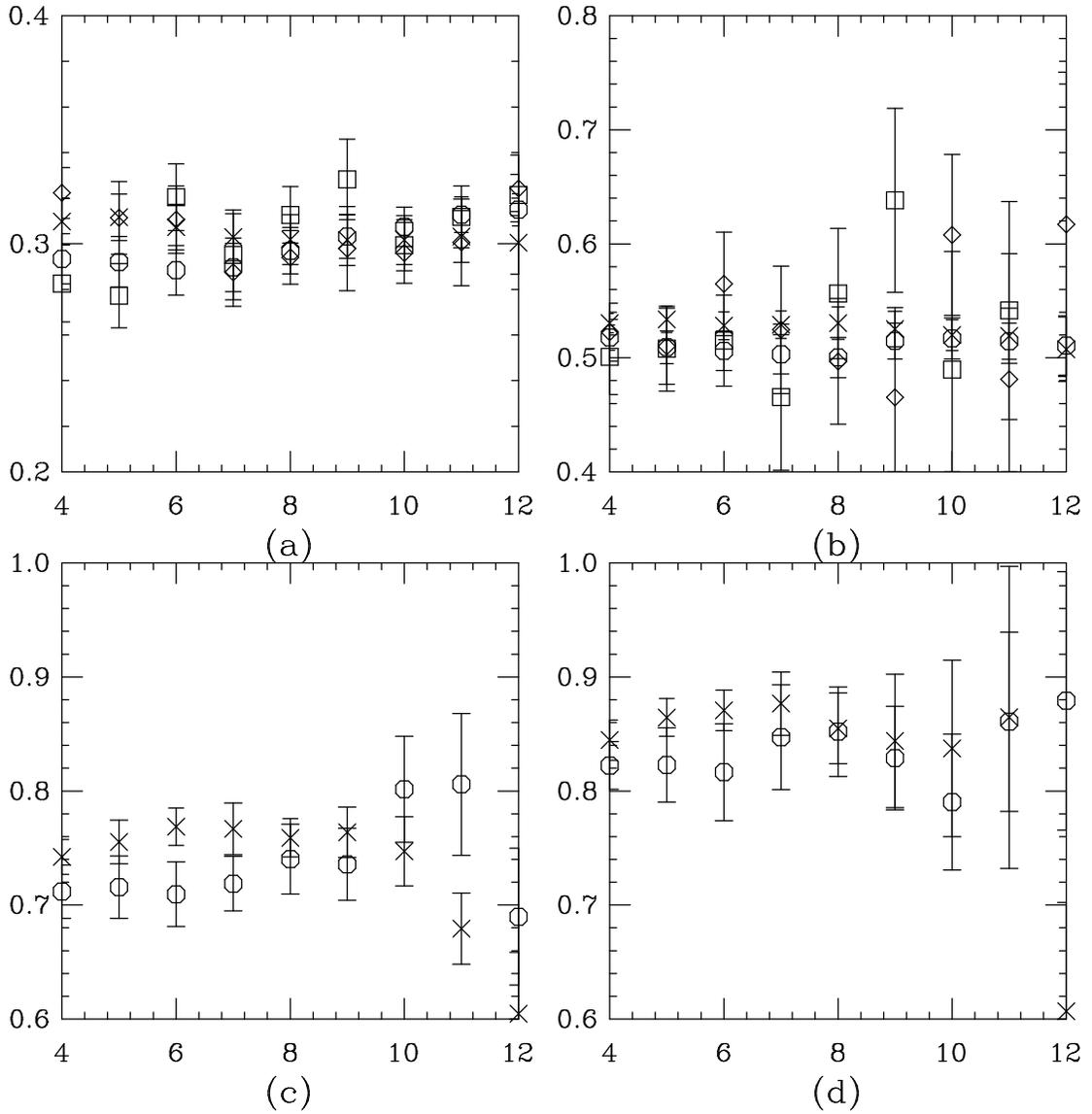

FIGURE 19

Effective mass fits to $\kappa = 0.1675$ data: (a) pion, (b) rho, (c) proton, and (d) delta. Data are labelled by type (WP or WW) and kind (1 or 2) by crosses (WP1), circles (WP2), diamonds (WW1) and squares (WW2).

from the pion mass squared, 0.1715(1) and from the quark mass, 0.1709(1). Of the four input kappas 0.1390, 0.1540, 0.1615, and 0.1670 only combinations of the last two are used in the fits (3 combinations of mass), since non-jackknife fits show that the other data are not linear in $\kappa$ or $1/\kappa$. This discrepancy is probably an artefact of heavy masses. The

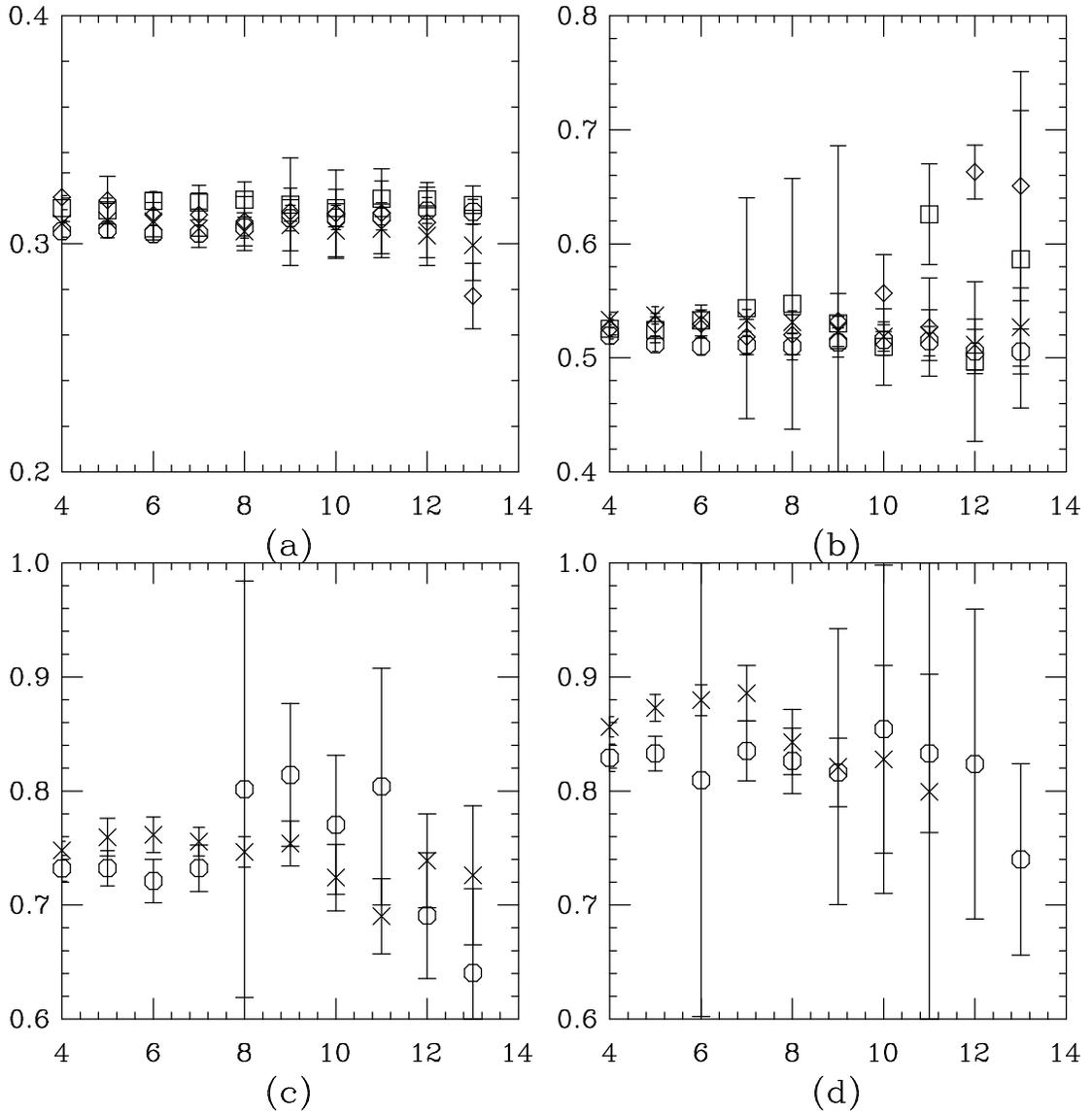

FIGURE 20

Fits from $t = D_{min}$ to 16 to $\kappa = 0.1675$ data: (a) pion, (b) rho, (c) proton, and (d) delta. Particles are labelled as in Fig. 19.

quark masses in lattice units are 0.044, 0.077, and 0.11 and the squared pion masses in lattice units are 0.21, 0.35, and 0.49. These are heavy masses compared to the ones we used in our previous work with staggered sea quarks and valence Wilson quarks, where the quark masses ranged from about 0.02 to 0.046 and the squared pion masses from 0.05 to 0.10 [18].

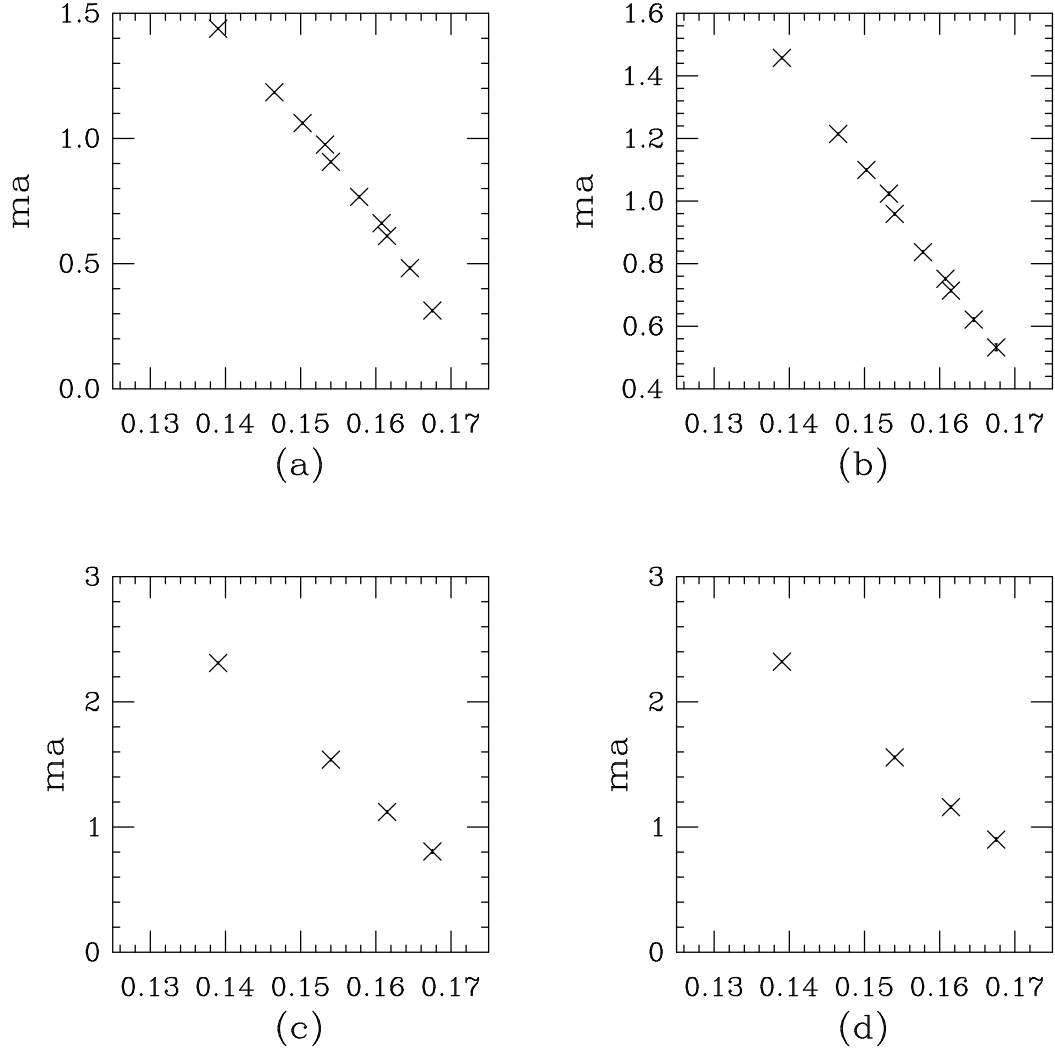

FIGURE 21

Best fit masses vs. average hopping parameter for $\kappa = 0.1675$ dynamical fermion data: (a) pion, (b) rho, (c) proton, and (d) delta.

As a test, we re-analyzed a subset of the data of Ref. [18] and picked out two kappas for which the masses are similar: squared pion masses of about 0.2, 0.3 and 0.41 gave $\kappa_c = 0.1604(1)$ while quark masses of 0.065, 0.09 and 0.13 or 0.035, 0.05 and 0.065 each gave $\kappa_c = 0.1608(1)$. The result from the lightest mass data was $\kappa_c = 0.1610(1)$.

As a consequence, we do not really know where the valence $\kappa_c$ is, for this value of the

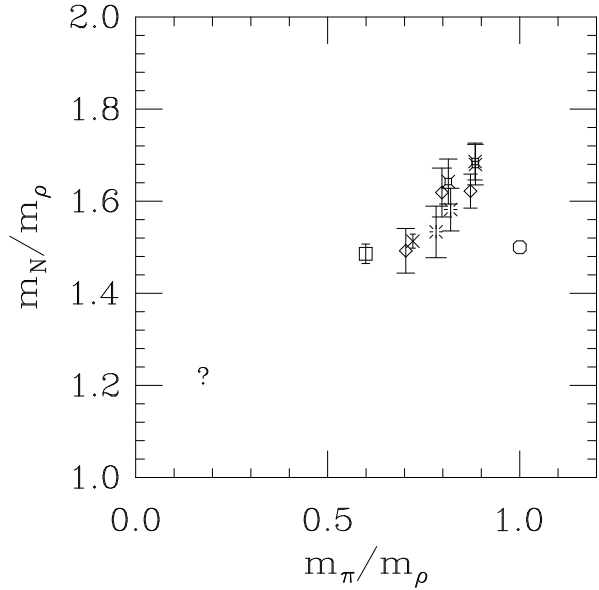

FIGURE 22

Edinburgh plot of our results. The square is our result at sea $\kappa = 0.1675$ and the cross is $\kappa = 0.1670$. The fancy squares, diamonds, and bursts are by Gupta, et. al. [23], at $\beta = 5.4$, 5.5, and 5.6 respectively. The circle and question mark show the expected values in the limit of infinite quark mass and from experiment.

sea quark mass. Not knowing $\kappa_c$ slightly affects masses and the lattice spacing since we don't know how far to extrapolate. However, our lightest spectroscopy is so far away from $\kappa_c$ that it makes no practical difference. Finally, the tadpole-normalized matrix elements need a factor $1 - 0.75\kappa/\kappa_c$ to convert from lattice to continuum. For $\kappa_c = 0.1709$ this factor is 0.390, 0.324, 0.291, 0.267 for our four quark masses, and for $\kappa_c = 0.1715$ it is 0.392, 0.326, 0.293, 0.270: i. e. again no practical difference.

At $\kappa = 0.1675$ the situation is better. The $\gamma_5$ and $\gamma_0\gamma_5$ pion masses extrapolate to 0.16970(7) and 0.16964(22) respectively, and the local and nonlocal axial current quark masses extrapolate to 0.16940(10) and 0.16933(9). The three quark masses are about 0.09, 0.05, and 0.02, and the three squared masses of the pions are about 0.37, 0.23, and 0.10.

Finally, the true $\kappa_c$ is obtained by extrapolating the square of the pion mass to zero and by extrapolating the quark mass. Both $\gamma_5$ and $\gamma_0\gamma_5$ pions extrapolated to a consistent value, and using both masses together gave $\kappa_c = 0.16794(2)$. The quark mass extracted from the local axial current gave $\kappa_c = 0.16795(4)$ while the quark mass extracted from the nonlocal axial current gave $\kappa_c = 0.16794(2)$.

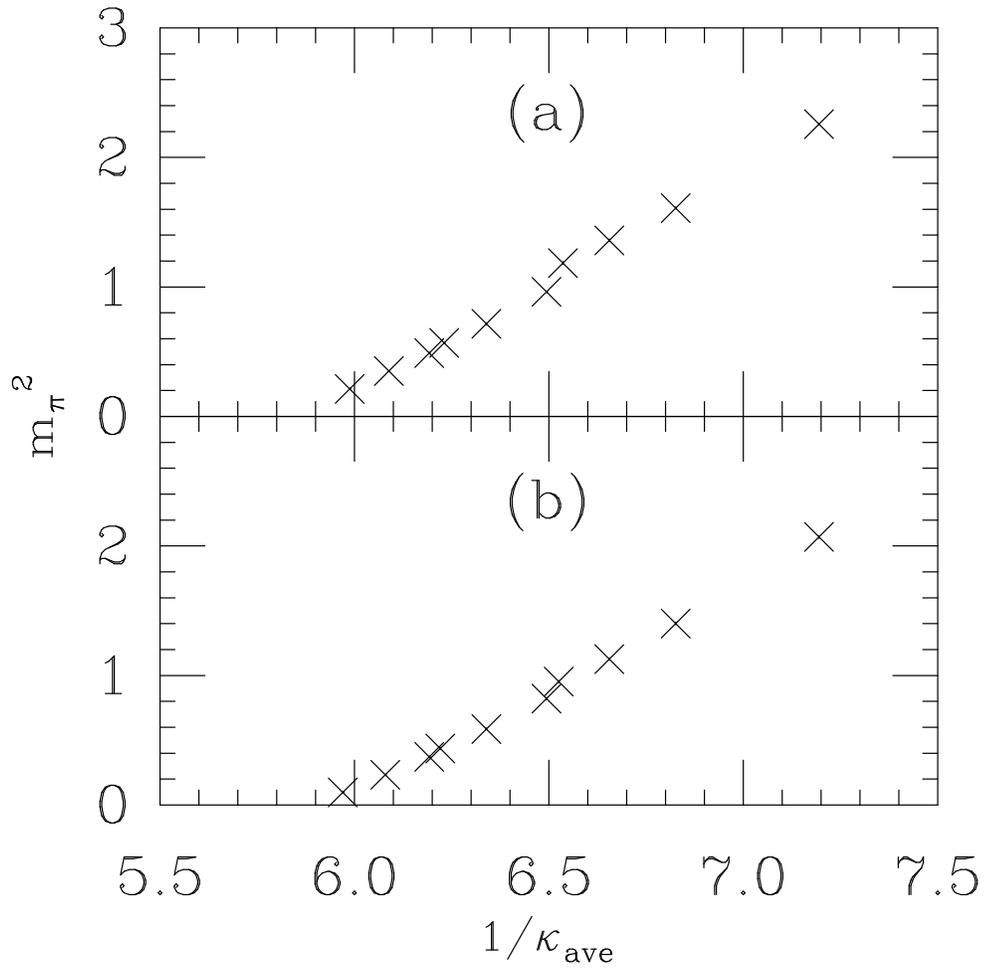

FIGURE 23

Square of pion mass vs. valence $1/\kappa$, for (a) $\kappa = 0.1670$ and (b) 0.1675 dynamical fermions.

Lattice Spacings

We can compute lattice spacings by extrapolating various masses to $\kappa_c$ and fixing the lattice spacing from them. Again, there are three possibilities: we can extrapolate in the valence hopping parameter at fixed sea quark hopping parameter, or we can extrapolate masses with degenerate sea and valence quark masses to $\kappa_c$. In all cases we have two quark masses which are light and so all hadron masses result from extrapolating linearly with two input masses.

At sea $\kappa = 0.1670$ the extrapolated rho, nucleon and delta lattice masses are $0.50(1)$, $0.71(1)$ and $0.84(2)$, giving inverse lattice spacings of $1540(36)$, $1324(20)$, and $1470(34)$ MeV, while at sea $\kappa = 0.1675$ the three lattice masses are $0.47(1)$, $0.70(2)$, and $0.82(2)$, giving inverse lattice spacings of $1638(35)$, $1340(30)$, and $1502(40)$ MeV. All errors are purely statistical.

A more sensible approach is to find the best-fit lattice spacing using all masses as input. If we do this we find that $1/a = 1415(15)$ at $\kappa = 0.1670$ so that $m_\rho = 707(7)$ MeV, $m_N = 1005(10)$ MeV and $m_\Delta = 1189(12)$ MeV. At $\kappa = 0.1675$ the lattice spacing is $1/a = 1532(21)$ MeV and the three masses are $719(10)$, $1072(15)$ and $1256(17)$ MeV.

For the true $\beta = 5.3$ lattice spacing we extrapolate the rho, nucleon, and delta to $\kappa_c$ and find lattice masses of $0.424(7)$, $0.593(17)$ and $0.775(27)$, which when compared to physical masses give inverse lattice spacings of $1816(30)$, $1585(45)$ and $1589(55)$ MeV. A common fit to all three masses gives $1/a = 1741(23)$ MeV and masses of $738(9)$, $1032(13)$, and $1349(18)$ MeV. Like quenched simulations at these values of the lattice spacing, the extrapolated hyperfine splittings are smaller than experiment.

We can also extract a lattice spacing from $f_\pi$ using tadpole-improved perturbation theory: from the local axial current the lattice $f_\pi$ is $0.066(2)$ for an inverse lattice spacing of $2000(61)$ MeV, while from the nonlocal axial current the corresponding numbers are $0.050(2)$ and $2640(100)$ MeV, respectively. At our $\beta$ value the nonlocal axial current is a bit smaller than the local current after inclusion of lattice to continuum renormalization factors. We return to this point in the next section.

## IV. MATRIX ELEMENTS

We calculated the same set of matrix elements on these lattices as we did in our earlier study of Wilson fermion matrix elements[18]: the decay constants of vector and pseudoscalar mesons and quark masses as extracted from the divergence of the axial current. The analysis is identical to that of Ref. 18; the reader is referred there for details. We remind the reader that one can perform either a "conventional" analysis, where the lattice to continuum fermion field renormalization is $\sqrt{2\kappa}$ or a tadpole-improved analysis[13], where the field renormalization is $\sqrt{1 - (3\kappa)/(4\kappa_c)}$ and the coefficients of $g^2$ in perturbative corrections to the operators are slightly modified through an all-orders resummation of tadpole diagrams.

Non-jackknifed matrix elements at $\kappa = 0.1670$ are computed by blocking five contiguous lattices together before performing correlated fits, and for $\kappa = 0.1675$ we blocked three lattices together. (We would like to have done more, but if we reduce the data set too far, the correlation matrix becomes singular.) For operators where a jackknife analysis is required, we performed a jackknife dropping sets of six contiguous blocked lattices at $\kappa = 0.1670$, and at $\kappa = 0.1675$ we blocked two contiguous lattices together, then performed a jackknife removing 10 successive blocked lattices from the ensemble.

### Vector Meson Decay Constant

We measured matrix elements of three vector current operators, the "local" vector current
$$V_\mu^l = \bar{\psi}\gamma_\mu\psi \tag{4.1}$$
the "nonlocal" current
$$V_\mu^{nl} = \frac{1}{2}(\bar{\psi}\gamma_\mu U_\mu \psi + h.c.) \tag{4.2}$$
and the conserved Wilson current
$$V_\mu^W = \frac{1}{2}(\bar{\psi}(U_\mu(\gamma_\mu - 1) + U_\mu^\dagger(\gamma_\mu + 1))\psi). \tag{4.3}$$

We extract the current matrix element from correlated fits to three parameters of two propagators with the appropriate operator as an interpolating field.

We quote our vector current matrix elements through the dimensionless parameter $f_V$
$$Z_V \langle V|V_\mu|0\rangle = \frac{1}{f_V} m_V^2 \epsilon_\mu. \tag{4.4}$$

The Wilson current is conserved but the other currents are multiplicatively renormalized. We measure these factors by doing a correlated fit to the Wilson current and to one of the nonconserved currents. This will extract the $(1 + A\alpha_V)$ part of $Z$. Our results are shown in Fig. 24. Tadpole-improved perturbation theory predicts that the $Z$ factors are $\kappa$ independent and equal to $(1 - 0.82\alpha_V)$ for the local current and $(1 - 1.00\alpha_V)$ for the nonlocal current [13]. In tadpole improved perturbation theory these numbers are expected to be 0.67 and 0.6 with $\alpha_v \simeq 0.4$. Our results show a five to ten per cent variation with valence $\kappa$ and a similar variation with sea quark mass. Note that our $Z$-factor for the nonlocal current is larger than for the local current; the reverse is true for the perturbative result.

In the Introduction we showed a comparison of the matrix element for the conserved vector current to data. We saw about a ten per cent variation with sea quark mass, with the lower number corresponding to lower sea quark mass. An extrapolation of the $\kappa = 0.1670$ and $0.1675$ $f_V$ to $\kappa_c$ gives $0.199(4)$ to be contrasted with 0.28 or 0.25 for the rho or omega. While one might expect that the matrix element $1/f_V$ for heavy quarks would be underestimated if the sea quark mass were too heavy [24], that argument cannot be extended to light quark systems. In any event, our results for $f_V$ resemble those of our earlier simulations with valence Wilson and dynamical staggered fermions [18].

## Quark Masses

We have already presented a determination of $\kappa_c$ from extrapolations of the quark mass from the current algebra relation

$$Z_A \nabla_\mu \cdot \langle \bar{\psi}\gamma_5\psi(0)\bar{\psi}\gamma_5\gamma_\mu\psi(x)\rangle = 2m_q \langle \bar{\psi}Z_P\gamma_5\psi(0)\bar{\psi}\gamma_5\psi(x)\rangle \qquad (4.5)$$

We have included the lattice-to-continuum Z-factors in the definition. We measured matrix elements of two axial current operators, the local current

$$A_0^{loc} = \bar{\psi}\gamma_0\gamma_5\psi \qquad (4.6)$$

and the nonlocal operator

$$A_0^{nl} = \frac{1}{2}(\bar{\psi}U_0\gamma_0\gamma_5\psi + h.c.). \qquad (4.7)$$

The ratio of the two currents is predicted to be [13] $Z_L/Z_{NL} = (1-0.31\alpha_V)/(1-0.9\alpha_V) \simeq$ 1.4 at $\alpha_v = 0.4$. Our measurement, shown in Fig. 25, is closer to about 1.2 and shows a twenty per cent variation with valence $\kappa$ with a small sea $\kappa$ dependence. This is remarkably good agreement with tadpole improved perturbation theory when one recalls that $\alpha_V = 0.4$.

We determine quark masses from both local and nonlocal axial currents. We show in Fig. 26 quark masses for all combinations of quarks as a function of $1/\kappa_{ave} - 1/\kappa_c$ where

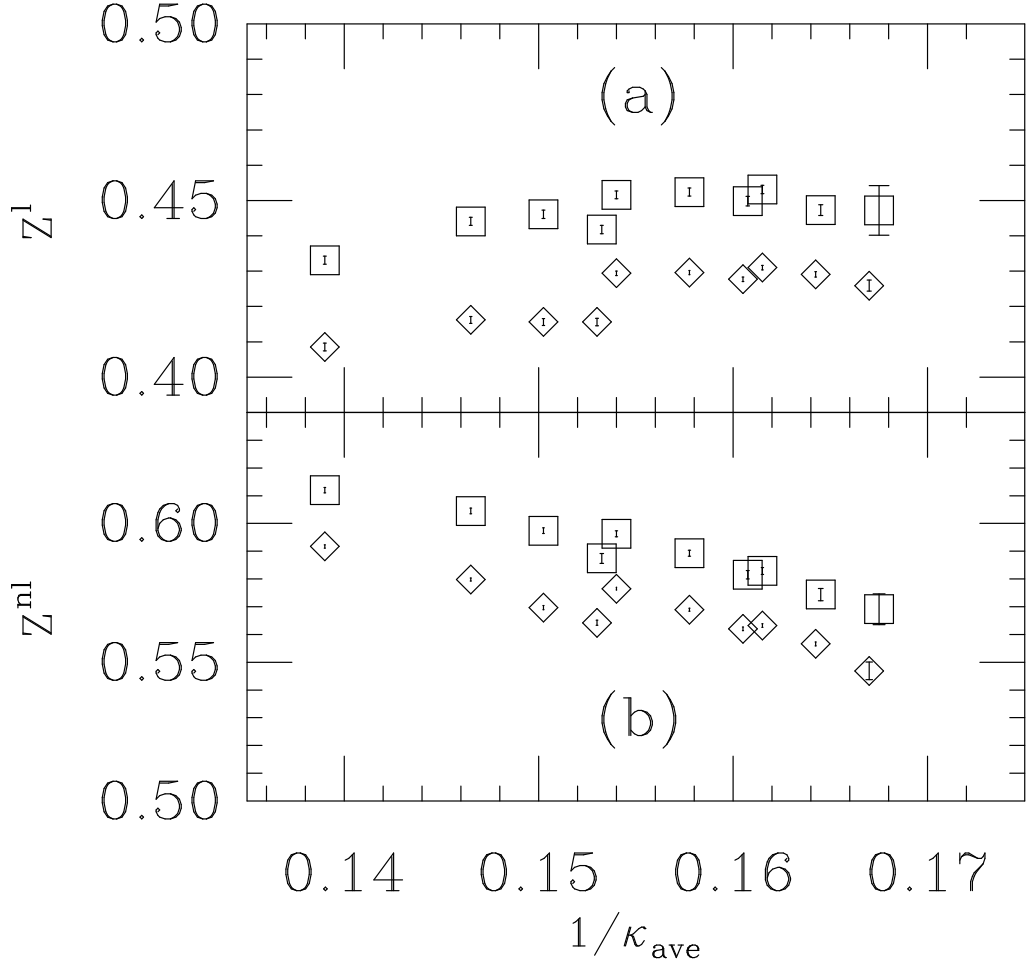

FIGURE 24

Ratios of renormalization factors for (a) local and (b) nonlocal vector currents to the conserved current. Results from simulations with sea quark mass $\kappa = 0.1675$ are shown in squares, and for sea quark mass $\kappa = 0.1670$ in diamonds.

$1/\kappa_{ave} = 0.5(1/\kappa_1 + 1/\kappa_2)$. We also show curves corresponding to

$$am_q = \frac{1}{2}(\frac{1}{\tilde{\kappa}} - 8) \qquad (4.8)$$

and

$$am_q = \log(\frac{1 - 6\tilde{\kappa}}{2\tilde{\kappa}}) \qquad (4.9)$$

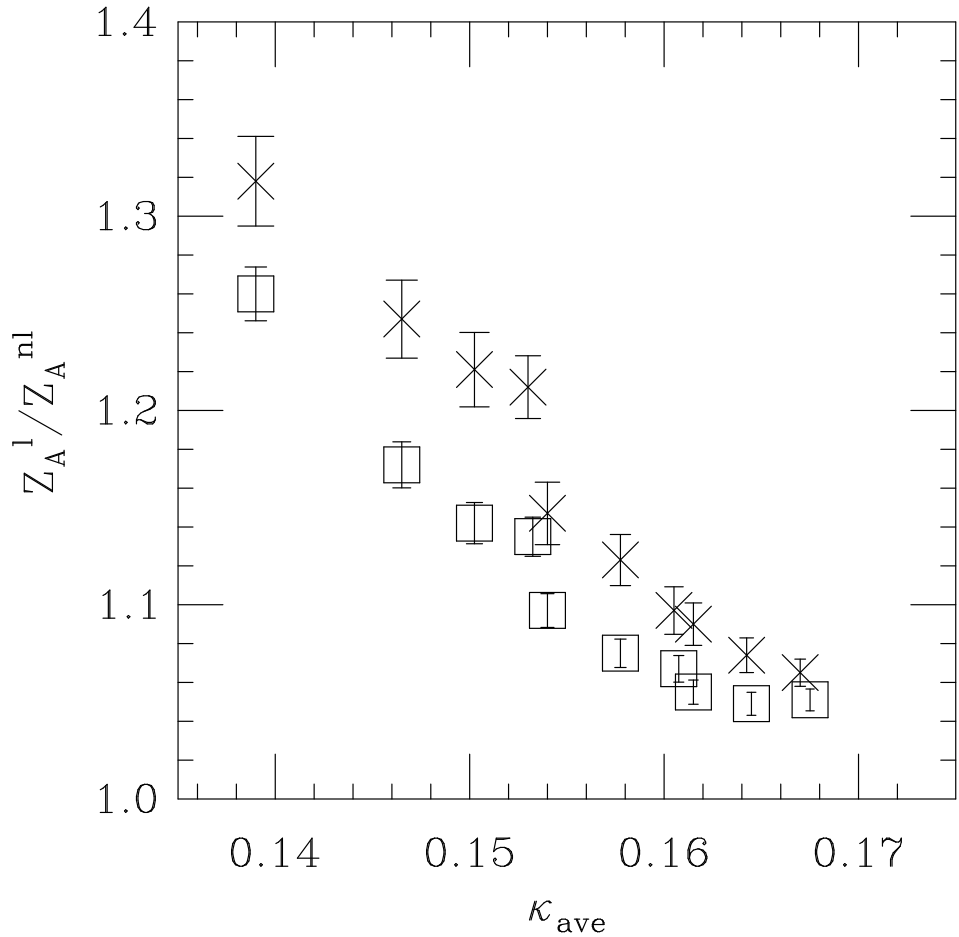

FIGURE 25

Ratios of renormalization factors of nonlocal to local axial currents. Results from simulations with sea quark $\kappa = 0.1675$ are shown in squares, and for sea quark mass $\kappa = 0.1670$ in crosses.

with $\tilde{\kappa} = \kappa_{ave}/(8\kappa_c)$. Either of these formulas reproduces the quark mass from the local axial current. The quark mass from the nonlocal axial current is a bit smaller, reflecting the fact that our observed $Z_L/Z_{NL}$ is a bit larger than the perturbative value.

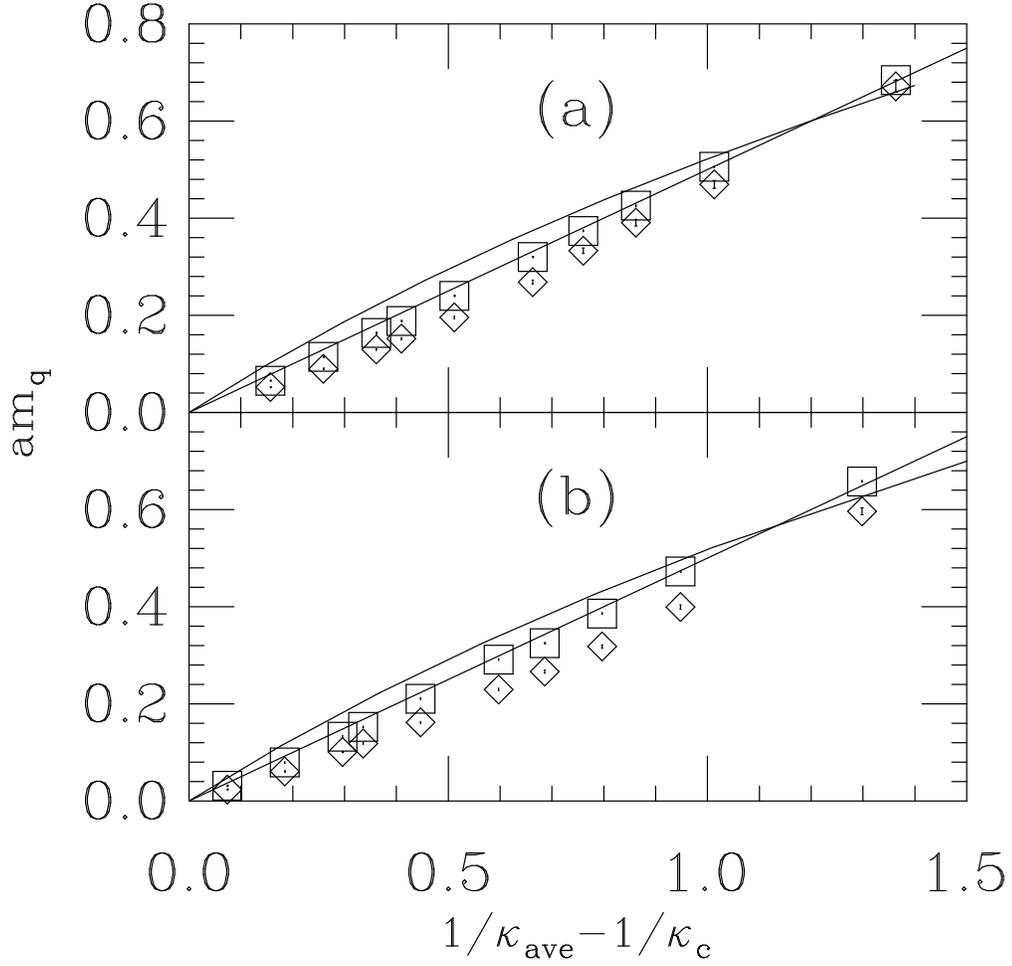

FIGURE 26

Quark mass from local (squares) and nonlocal (diamonds) axial currents, as a function of $1/\kappa_{ave} - 1/\kappa_c$, for dynamical $\kappa = 0.1670$ (a) and 0.1675 (b). The curves are the simple quark mass and tadpole quark mass described in the text.

Pseudoscalar Decay Constant

Our extraction of pseudoscalar decay constants parallels other recent quenched analyses of these quantities. Note however that our lattice spacing is considerably larger than what is used in contemporary quenched simulations. This introduces an unknown systematic effect on our results. Ref. [25] shows a ten per cent drop in the ratio $f_K/f_\pi$ in going from quenched simulations at $1/a \simeq 1.2$ to $1/a \simeq 3.1$ GeV, and one could easily imagine that the effect is bigger for mesons containing heavier quarks. Indeed our numbers are somewhat greater than quenched results from smaller lattice spacing analyzed in a similar way to ours [25]. Thus we do NOT regard the results we will present as serious QCD predictions for pseudoscalar decay constants. The one result we wish to examine is the degree of dependence of matrix elements on the sea quark mass, which might be used to infer the degree to which quenched simulations might be modified by the inclusion of virtual $q\bar{q}$ pairs.

We compute matrix elements of the axial currents of Eqns. (4.6) and (4.7) and extract a "raw" decay constant $f_P$ from the definition $\langle 0|A_0|P\rangle = m_P f_P$. In our conventions the experimental $f_\pi = 132$ MeV. We determined $f_P$ for all combinations of quark mass. In order to arrive at physical numbers we then carried out the following steps:

1. We extrapolated heavy meson decay constants to zero light quark mass by a linear extrapolation in the light quark hopping parameter to $\kappa_c$, using the two lightest quark hopping parameters in each data set ($\kappa = 0.1615$ and either 0.1670 or 0.1675). This extrapolation included the $\kappa$-dependent field normalization and appropriate Z-factor. For tadpole-improved matrix elements this includes the $\sqrt{1-6\tilde{\kappa}}$ factor which allows Wilson fermions to interpolate to the infinite mass limit.

2. Our heavy quarks have masses which are large compared to an inverse lattice spacing. In this limit the dispersion relation for free Wilson fermions is

$$E(\vec{k}) = m_1 + \frac{\vec{k}^2}{2m_2} + \ldots \tag{4.10}$$

where $m_1$ is given by Eqn. (4.9) and

$$am_2 = \frac{\exp(am_1)\sinh(am_1)}{1 + \sinh(am_1)}. \tag{4.11}$$

Kronfeld [26], Mackenzie [27] and Bernard, Labrenz, and Soni [25] argue that the appropriate quark mass at which the matrix element is measured is not $m_1$ but $m_2$ since it enters in the kinetic energy while $m_1$ is just an overall additive constant. Their analysis suggests that we correct for this error by adjusting the meson mass

$$aM \to aM' = aM + (am_2 - am_1). \tag{4.12}$$

This is a shift of no more than 0.125 at $\kappa = 0.1390$.

The "raw" $f_P$ data (no kappa factors, no Z-factors) for the local operator for sea $\kappa = 0.1670$ and 0.1675 are shown in Tables XII-XIII. Fig. 27 displays plots of $af_P\sqrt{aM_P}$ vs $1/aM_P$ for heavy-light systems, including the extrapolated zero light quark mass points.

Finally, we convert to physical units by fixing the lattice spacing from $f_\pi$. As we have already remarked, for the local axial current this number is close to other determinations of the lattice spacing we can make from spectroscopy. For the nonlocal axial current the lattice spacing is fairly sensitive to the lattice-to-continuum regularization convention; in two extreme cases $1/a = 1040$ MeV from the $\kappa = 0.1670$ conventional renormalization to $1/a = 2465$ MeV from the $\kappa = 0.1675$ tadpole-improved renormalization.

Let us consider a few examples of determinations of decay constants from specific subsets of the data, focussing on the tadpole improved numbers.

The meson mass at the heaviest valence quark mass from the local current is accidentally close to the D meson mass. The $\kappa = 0.1675$ number for $f_D$ from the local current at an inverse lattice spacing of $1/a = 1935$ MeV is 270(7) MeV and scales linearly with $1/a$. At $\kappa = 0.1670$ we have 290(5) MeV, with $1/a = 1618$ MeV. The uncertainties quoted are purely statistical.

We can extract a prediction for $f_{D_s}$ using the $\kappa = 0.1390 - 0.1670$ decay constant, since the 0.1670 hadrons have the closest pseudoscalar-vector ratio to strange quarks (compare Fig. 3). With $1/a = 1619$ MeV, the tadpole-improved local axial current $f_{D_s}$ is 330(7) MeV. The nonlocal current gives 358(6) MeV. Again, the errors are purely statistical and does not include any uncertainty due to lattice spacing. The number has recently been determined by two experiments to be $232 \pm 45 \pm 20 \pm 48$ MeV [28] or $344 \pm 37 \pm 52 \pm 42$ MeV[29].

We attempted to fit these data to $f\sqrt{M} = A + B/M + C/M^2$ and then to extrapolate to the physical D and B masses. Two parameter fits had $\chi^2$ in the range 8-60. We have three decay constants per operator/Z-factor combination and so three parameter fits will have $\chi^2 = 0$. The fit values of $f_D$ are stable under changing from two to three parameters within about 20 MeV.

We showed calculations of the decay constant with conventional normalization in Fig. 5. They are lower than the tadpole improved predictions. This is not surprising since the relative normalization of the two schemes is $\sqrt{(1-6\tilde\kappa)/(2\kappa)} \simeq 1.2$ at the heaviest quark mass. Our data look very much like results from many other quenched simulations over a wide range of lattice spacing values. Two parameter fits as described above had $\chi^2$ in the range 3-10.

How can we assign a real uncertainty to these numbers? The main sources of error are systematic. We believe that our determination of the lattice spacing has an uncertainty of fifteen per cent, or 40 MeV at $f_D = 250$ MeV. Differences in the final result from choice of operator (local versus nonlocal current) are in the range 10 to 20 MeV and will be quoted below. A big systematic is the choice of lattice to continuum renormalization. We will quote numbers from both conventions since the choice of a particular one has not been definitely settled in the literature. Other uncertainties such as the precise value of $\alpha_V$ are small, order 5 MeV.

Thus we calculate

$$f_D = 250(5) \pm 40 \pm 20 \pm 5 \ MeV \tag{4.13}$$

$$f_{D_s} = 345(5) \pm 45 \pm 15 \pm 5 \ MeV \tag{4.14}$$

$$f_B = 200(10) \pm 40 \pm 15 \pm 5 \ MeV \tag{4.15}$$

from tadpole improved perturbation theory, while

$$f_D = 175(5) \pm 40 \pm 20 \pm 5 \ MeV \tag{4.16}$$

$$f_{D_s} = 220(5) \pm 40 \pm 15 \pm 5 \ MeV \tag{4.17}$$

$$f_B = 125(5) \pm 40 \pm 20 \pm 5 \ MeV \tag{4.18}$$

from a conventional perturbative calculation. The error in parentheses is statistical (including extrapolation) and the three other uncertainties represent lattice spacing, choice of operator, and $\alpha_V$ uncertainty. We have included no uncertainty associated with the sea quark mass; it is lumped in with the statistical/extrapolation uncertainty. We do not see any observable effects of different sea quark masses. These calculations are a bit higher than quenched calculations done at smaller values of the lattice spacing [25,30], though except for $f_{D_s}$, not outside error bars. It may be that the effect of dynamical fermions is to push up the matrix elements but it is also possible (and more likely, in our opinion) that the large lattice spacing induces a systematic shift upwards in the decay constant, especially for the tadpole-improved matrix elements.

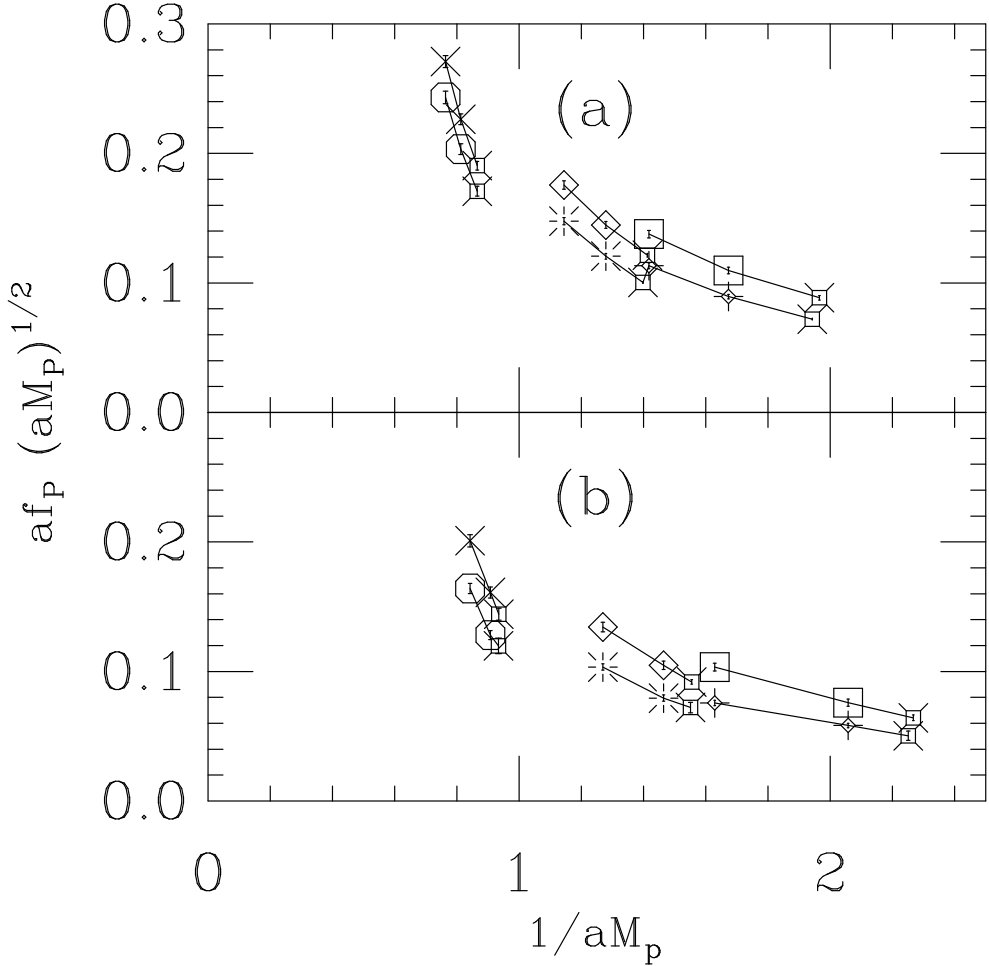

FIGURE 27

Lattice pseudoscalar decay constants, $af_P\sqrt{aM_P}$ vs. $1/aM_P$ for sea quark $\kappa = 0.1670$ (a) and $0.1675$ (b). Data are labeled with crosses, diamonds, and squares for heavy quark $\kappa = 0.1390, 0.1540$, and $0.1615$, for the local operator, and octagons, bursts, and fancy diamonds for the nonlocal operator. The fancy squares show the extrapolation to $\kappa_c$.

## V. CONCLUSIONS

We carried out a simulation of QCD with two degenerate flavors of reasonably light sea quarks at $6/g^2 = 5.3$, corresponding to a lattice spacing of roughly $a = 0.12$ to $0.13$ fm.

The lightest pseudoscalar- vector mass ratio we achieved was about 0.6. The spectroscopy appears qualitatively similar to quenched simulations at large lattice spacing; effects of sea quarks on masses are small.

We studied simple matrix elements and saw that the influence of sea quarks on physical observables was small when those observables are expressed as a function of other physical observables. This is a "simulational justification" of the quenched approximation, though admittedly for big sea quark masses. Although the coupling constant is large, tadpole improved perturbation theory does a good job of predicting $\kappa_c$ and the ratios of renormalization factors among different lattice choices for currents. Our results are not too different from those from quenched simulations at large values of the lattice spacing.

However, these results are not yet satisfactory for doing precision calculations in QCD. We need to push to smaller values of the sea quark mass. We also need either to push to smaller values of the lattice spacing or to continue to develop techniques which allow one to carry out simulations at large lattice spacing which have smaller intrinsic discretization systematics than present simulations do.

## ACKNOWLEDGMENTS

This work was supported by the U. S. Department of Energy under contracts DE–FG05–92ER–40742, DE–FG02–85ER–40213, DE–FG02–91ER–40672, DE–AC02–84ER–40125, W-31-109-ENG-38, and by the National Science Foundation under grants NSF-PHY87-01775, NSF-PHY89-04035 and NSF-PHY91-16964. The computations were carried out at the Florida State University Supercomputer Computations Research Institute which is partially funded by the U.S. Department of Energy through Contract No. DE-FC05-85ER250000. T. D. would like to thank J. Labrenz for discussions about Ref. [25] and for providing data, and P. Mackenzie for discussions about Ref. [13]. We thank T. Kitchens and J. Mandula for their continuing support and encouragement.

Table I. Data set for Wilson fermions. $\Delta T$ is the simulation time interval between stored lattices.

| $\kappa$ | source | number of lattices | $\Delta T$ |
|---|---|---|---|
| 0.1670 | wall | 484 | 5 |
|  | shell | 241 | 10 |
| 0.1675 | wall | 417 | 3 |
|  | shell | 141 | 9 |

Table II. Pseudoscalar fits — dynamical quark hopping parameter $\kappa = 0.1670$. All fits are to a single exponential. In the following tables, the label "kind" numbers the hopping parameters of the quarks which make up the hadron, 1 for $\kappa = 0.1390$, 2 for 0.1540, 3 for 0.1615, and 4 for 0.1670 or 0.1675. The label "WP" designates the wall-point correlator. The label $\kappa$ is the average hopping parameter of the constituents.

| Kind | $\kappa$ | $D_{min}$ | $D_{max}$ | mass | $\chi^2/dof$ | C.L. |
|---|---|---|---|---|---|---|
| 1 1 | 0.1390 | 5 | 16 | 1.502( 2) | 83.328/10 | 0.000 |
| 2 1 | 0.1465 | 4 | 16 | 1.268( 2) | 139.187/11 | 0.000 |
| 2 2 | 0.1540 | 4 | 16 | 0.981( 2) | 16.621/11 | 0.120 |
| 3 1 | 0.1502 | 11 | 16 | 1.166( 3) | 167.837/4 | 0.000 |
| 3 2 | 0.1578 | 9 | 16 | 0.845( 2) | 7.849/6 | 0.249 |
| 3 3 | 0.1615 | 11 | 16 | 0.699( 3) | 3.262/4 | 0.515 |
| 4 1 | 0.1530 | 11 | 16 | 1.088( 3) | 143.280/4 | 0.000 |
| 4 2 | 0.1605 | 11 | 16 | 0.753( 3) | 3.930/4 | 0.416 |
| 4 3 | 0.1643 | 11 | 16 | 0.591( 3) | 4.869/4 | 0.301 |
| 4 4 | 0.1670 | 11 | 16 | 0.462( 3) | 7.719/4 | 0.102 |
| WP | 0.1670 | 8 | 16 | 0.454( 2) | 4.909/7 | 0.671 |

Table III. Vector meson fits — dynamical quark hopping parameter $\kappa = 0.1670$, as in Table II.

| Kind | $\kappa$ | $D_{min}$ | $D_{max}$ | mass | $\chi^2/dof$ | C.L. |
|---|---|---|---|---|---|---|
| 1 1 | 0.1390 | 7 | 16 | 1.522( 2) | 48.999/8 | 0.000 |
| 2 1 | 0.1465 | 4 | 16 | 1.293( 2) | 73.095/11 | 0.000 |
| 2 2 | 0.1540 | 4 | 16 | 1.038( 2) | 21.854/11 | 0.026 |
| 3 1 | 0.1502 | 4 | 16 | 1.190( 2) | 89.072/11 | 0.000 |
| 3 2 | 0.1578 | 4 | 16 | 0.923( 2) | 17.751/11 | 0.088 |
| 3 3 | 0.1615 | 4 | 16 | 0.802( 2) | 9.931/11 | 0.537 |
| 4 1 | 0.1530 | 4 | 16 | 1.112( 2) | 56.072/11 | 0.000 |
| 4 2 | 0.1605 | 4 | 16 | 0.843( 2) | 8.596/11 | 0.659 |
| 4 3 | 0.1643 | 4 | 16 | 0.717( 2) | 8.758/11 | 0.644 |
| 4 4 | 0.1670 | 8 | 16 | 0.636( 5) | 6.674/7 | 0.464 |
| WP | 0.1670 | 7 | 16 | 0.635( 2) | 6.670/8 | 0.573 |

Table IV. Nucleon fits — dynamical quark hopping parameter $\kappa = 0.1670$, as in Table II.

| Kind | $\kappa$ | $D_{min}$ | $D_{max}$ | mass | $\chi^2/dof$ | C.L. |
|---|---|---|---|---|---|---|
| 1 | 0.1390 | 4 | 16 | 2.386( 4) | 6.288/11 | 0.853 |
| 2 | 0.1540 | 4 | 16 | 1.666( 6) | 18.110/11 | 0.079 |
| 3 | 0.1615 | 7 | 16 | 1.266( 8) | 7.241/8 | 0.511 |
| 4 | 0.1670 | 10 | 16 | 0.987(18) | 4.693/5 | 0.454 |
| WP | 0.1670 | 6 | 16 | 0.962( 4) | 9.647/9 | 0.380 |

Table V. Delta fits — dynamical quark hopping parameter $\kappa = 0.1670$, as in Table II.

| Kind | $\kappa$ | $D_{min}$ | $D_{max}$ | mass | $\chi^2/dof$ | C.L. |
|---|---|---|---|---|---|---|
| 1 | 0.1390 | 4 | 16 | 2.392( 4) | 6.397/11 | 0.846 |
| 2 | 0.1540 | 4 | 16 | 1.686( 6) | 16.415/11 | 0.126 |
| 3 | 0.1615 | 5 | 16 | 1.326( 8) | 13.941/10 | 0.176 |
| 4 | 0.1670 | 4 | 16 | 1.060( 9) | 7.645/11 | 0.745 |
| WP | 0.1670 | 6 | 16 | 1.044( 6) | 11.072/9 | 0.271 |

Table VI. Pseudoscalar fits — dynamical quark hopping parameter $\kappa = 0.1675$, as in Table II.

| Kind | $\kappa$ | $D_{min}$ | $D_{max}$ | mass | $\chi^2/dof$ | C.L. |
|---|---|---|---|---|---|---|
| 1 1 | 0.1390 | 10 | 16 | 1.438( 2) | 4.123/5 | 0.532 |
| 2 1 | 0.1465 | 10 | 16 | 1.184( 3) | 3.559/5 | 0.614 |
| 2 2 | 0.1540 | 10 | 16 | 0.906( 3) | 3.178/5 | 0.673 |
| 3 1 | 0.1502 | 10 | 16 | 1.061( 3) | 2.848/5 | 0.723 |
| 3 2 | 0.1578 | 10 | 16 | 0.766( 3) | 3.509/5 | 0.622 |
| 3 3 | 0.1615 | 9 | 16 | 0.610( 3) | 4.743/6 | 0.577 |
| 4 1 | 0.1532 | 9 | 16 | 0.976( 3) | 2.994/6 | 0.810 |
| 4 2 | 0.1608 | 9 | 16 | 0.661( 3) | 4.245/6 | 0.644 |
| 4 3 | 0.1645 | 9 | 16 | 0.482( 4) | 3.926/6 | 0.687 |
| 4 4 | 0.1675 | 7 | 16 | 0.312( 4) | 8.820/8 | 0.358 |
| WP | 0.1675 | 9 | 16 | 0.309( 7) | 5.365/6 | 0.498 |

Table VII. Vector meson fits — dynamical quark hopping parameter $\kappa = 0.1675$, as in Table II.

| Kind | $\kappa$ | $D_{min}$ | $D_{max}$ | mass | $\chi^2/dof$ | C.L. |
|---|---|---|---|---|---|---|
| 1 1 | 0.1390 | 10 | 16 | 1.457( 3) | 3.580/5 | 0.611 |
| 2 1 | 0.1465 | 10 | 16 | 1.214( 3) | 5.098/5 | 0.404 |
| 2 2 | 0.1540 | 10 | 16 | 0.959( 4) | 8.162/5 | 0.148 |
| 3 1 | 0.1502 | 10 | 16 | 1.099( 3) | 7.169/5 | 0.208 |
| 3 2 | 0.1578 | 10 | 16 | 0.837( 4) | 9.032/5 | 0.108 |
| 3 3 | 0.1615 | 9 | 16 | 0.714( 4) | 9.556/6 | 0.145 |
| 4 1 | 0.1532 | 9 | 16 | 1.023( 5) | 6.171/6 | 0.404 |
| 4 2 | 0.1608 | 9 | 16 | 0.751( 4) | 4.682/6 | 0.585 |
| 4 3 | 0.1645 | 9 | 16 | 0.621( 6) | 5.606/6 | 0.469 |
| 4 4 | 0.1675 | 7 | 16 | 0.532(10) | 19.384/8 | 0.013 |
| WP | 0.1675 | 4 | 16 | 0.523( 4) | 7.569/11 | 0.751 |

Table VIII. Nucleon fits — dynamical quark hopping parameter $\kappa = 0.1675$, as in Table II.

| Kind | $\kappa$ | $D_{min}$ | $D_{max}$ | mass | $\chi^2/dof$ | C.L. |
|---|---|---|---|---|---|---|
| 1 | 0.1390 | 10 | 16 | 2.309( 7) | 2.019/5 | 0.847 |
| 2 | 0.1540 | 10 | 16 | 1.537( 9) | 2.010/5 | 0.848 |
| 3 | 0.1615 | 10 | 16 | 1.121(10) | 0.333/5 | 0.997 |
| 4 | 0.1675 | 5 | 16 | 0.804(13) | 5.523/10 | 0.854 |
| WP | 0.1675 | 6 | 16 | 0.766( 9) | 18.046/9 | 0.035 |

Table IX. Delta fits — dynamical quark hopping parameter $\kappa = 0.1675$, as in Table II.

| Kind | $\kappa$ | $D_{min}$ | $D_{max}$ | mass | $\chi^2/dof$ | C.L. |
|---|---|---|---|---|---|---|
| 1 | 0.1390 | 10 | 16 | 2.319( 7) | 3.089/5 | 0.686 |
| 2 | 0.1540 | 11 | 16 | 1.557(10) | 1.807/4 | 0.771 |
| 3 | 0.1615 | 10 | 16 | 1.159(10) | 0.697/5 | 0.983 |
| 4 | 0.1675 | 4 | 16 | 0.901(14) | 6.605/11 | 0.830 |
| WP | 0.1390 | 7 | 16 | 0.891(15) | 28.742/8 | 0.000 |

Table X. Mass ratios for Wilson fermions, $\kappa = 0.1670$.

| Kind | $\kappa$ | $D_{min}$ | $D_{max}$ | ratio | $\chi^2/dof$ | C.L. |
|---|---|---|---|---|---|---|
| $m_\pi/m_\rho$ | | | | | | |
| 1 | 0.1390 | 5 | 16 | 0.9870( 4) | 239.500/20 | 0.000 |
| 2 | 0.1540 | 7 | 16 | 0.9471( 5) | 45.880/16 | 0.000 |
| 3 | 0.1615 | 11 | 16 | 0.868( 2) | 6.530/8 | 0.367 |
| 4 | 0.1670 | 11 | 16 | 0.715( 8) | 10.510/8 | 0.105 |
| WP | 0.1670 | 10 | 16 | 0.722( 4) | 17.650/10 | 0.024 |
| $m_N/m_\rho$ | | | | | | |
| 1 | 0.1390 | 7 | 16 | 1.585( 2) | 78.180/16 | 0.000 |
| 2 | 0.1540 | 7 | 16 | 1.599( 4) | 40.080/16 | 0.000 |
| 3 | 0.1615 | 7 | 16 | 1.588( 8) | 12.890/16 | 0.535 |
| 4 | 0.1670 | 11 | 16 | 1.536(38) | 7.778/8 | 0.255 |
| WP | 0.1670 | 9 | 16 | 1.513(15) | 13.510/12 | 0.197 |

Table XI. Mass ratios for Wilson fermions, $\kappa = 0.1675$.

| Kind | $\kappa$ | $D_{min}$ | $D_{max}$ | ratio | $\chi^2/dof$ | C.L. |
|---|---|---|---|---|---|---|
| $m_\pi/m_\rho$ | | | | | | |
| 1 | 0.1390 | 10 | 16 | 0.987( 1) | 5.560/10 | 0.696 |
| 2 | 0.1540 | 11 | 16 | 0.945(23) | 13.900/8 | 0.031 |
| 3 | 0.1615 | 10 | 16 | 0.858(12) | 15.430/10 | 0.051 |
| 4 | 0.1675 | 8 | 16 | 0.577(16) | 25.920/14 | 0.011 |
| WP | 0.1675 | 9 | 16 | 0.599( 8) | 8.022/12 | 0.627 |
| $m_N/m_\rho$ | | | | | | |
| 1 | 0.1390 | 11 | 16 | 1.584(15) | 8.527/8 | 0.202 |
| 2 | 0.1540 | 10 | 16 | 1.601( 5) | 10.180/10 | 0.253 |
| 3 | 0.1615 | 10 | 16 | 1.585(13) | 11.100/10 | 0.196 |
| 4 | 0.1675 | 8 | 16 | 1.438(41) | 19.790/14 | 0.071 |
| WP | 0.1675 | 8 | 16 | 1.486(21) | 23.810/14 | 0.022 |

Table XII. Matrix elements of the local axial current from simulations with sea $\kappa = 0.1670$, with no Z-factors or lattice-to-continuum $\kappa$ renormalization.

| Kind | $\kappa$ | $D_{min}$ | $D_{max}$ | $f_P^L$ | $\chi^2/dof$ | C.L. |
|---|---|---|---|---|---|---|
| 1 1 | 0.1390 | 9 | 16 | 0.970(13) | 56.070/11 | 0.000 |
| 2 1 | 0.1465 | 10 | 16 | 0.869(14) | 71.220/9 | 0.000 |
| 2 2 | 0.1540 | 10 | 16 | 0.759(13) | 18.910/9 | 0.026 |
| 3 1 | 0.1502 | 10 | 16 | 0.797(13) | 83.490/9 | 0.000 |
| 3 2 | 0.1578 | 10 | 16 | 0.692(12) | 18.050/9 | 0.035 |
| 3 3 | 0.1615 | 10 | 16 | 0.637(12) | 19.530/9 | 0.021 |
| 4 1 | 0.1530 | 10 | 16 | 0.716(13) | 64.700/9 | 0.000 |
| 4 2 | 0.1605 | 10 | 16 | 0.630(11) | 15.890/9 | 0.069 |
| 4 3 | 0.1643 | 10 | 16 | 0.575(13) | 20.580/9 | 0.015 |
| 4 4 | 0.1670 | 9 | 16 | 0.490( 9) | 30.500/11 | 0.001 |

Table XIII. As Table XII, but $\kappa = 0.1675$.

| Kind | $\kappa$ | $D_{min}$ | $D_{max}$ | $f_P^L$ | $\chi^2/dof$ | C.L. |
|---|---|---|---|---|---|---|
| 1 1 | 0.1390 | 10 | 16 | 0.829(15) | 5.361/9 | 0.802 |
| 2 1 | 0.1465 | 10 | 16 | 0.727(15) | 6.641/9 | 0.674 |
| 2 2 | 0.1540 | 10 | 16 | 0.646(15) | 8.842/9 | 0.452 |
| 3 1 | 0.1502 | 10 | 16 | 0.634(14) | 8.923/9 | 0.444 |
| 3 2 | 0.1578 | 10 | 16 | 0.572(16) | 13.600/9 | 0.137 |
| 3 3 | 0.1615 | 9 | 16 | 0.527(15) | 19.800/11 | 0.048 |
| 4 1 | 0.1532 | 9 | 16 | 0.554(14) | 23.310/11 | 0.016 |
| 4 2 | 0.1608 | 9 | 16 | 0.504(15) | 21.530/11 | 0.028 |
| 4 3 | 0.1645 | 9 | 16 | 0.456(16) | 25.650/11 | 0.007 |
| 4 4 | 0.1675 | 9 | 16 | 0.388(20) | 23.620/11 | 0.014 |